\documentclass[aps,prl,twocolumn,groupedaddress]{revtex4-1}
\usepackage{graphicx}
\usepackage{caption}
\DeclareCaptionLabelSeparator{dot}{. }
\makeatletter
\def\justified{
	\let\\\@normalcr
	\@rightskip\z@skip \rightskip\@rightskip
	\leftskip\z@skip
	\parindent 0em\relax
	\setlength{\parfillskip}{0pt plus 1fil}}
\DeclareCaptionJustification{justified}{\justified}
\usepackage{subcaption}
\usepackage{amsmath}
\usepackage{braket}
\usepackage{nicefrac}
\usepackage{units}
\usepackage{ragged2e}
\usepackage[colorlinks,urlcolor=blue ,citecolor=blue ,linkcolor=blue ]{hyperref}
\usepackage{xcolor}
\usepackage{nicefrac} 

\newcommand{\Er}{{^{167}}{\rm Er}}
\newcommand{\nine}{{|\!\downarrow\,\rangle}}
\newcommand{\seven}{{|\!\uparrow\,\rangle}}
\newcommand{\five}{|\text{--}15/2\rangle}

\newcommand{\up}{\uparrow}
\newcommand{\down}{\downarrow}
\newcommand{\HC}{{\mathrm{ODT}_{1570}}}
\newcommand{\mopa}{{\mathrm{ODT}_{1064}}}

\begin{document}
	
	
	\title{Realization of a Strongly Interacting Fermi Gas of Dipolar Atoms}
	
	\author{S. Baier$^{1}$, D. Petter$^{1}$, J. H. Becher$^{1,\dagger}$, A. Patscheider$^{1,2}$, G. Natale$^{1}$, L. Chomaz$^1$, M. J. Mark$^{1,2}$, F. Ferlaino$^{1,2,*}$}
	
	\affiliation{%
		$^{1}$Institut f\"ur Experimentalphysik,Universit\"at Innsbruck, Technikerstra{\ss}e 25, 6020 Innsbruck, Austria\\
		$^{2}$Institut f\"ur Quantenoptik und Quanteninformation,\"Osterreichische Akademie der Wissenschaften, 6020 Innsbruck, Austria\\
	}

	
	\date{\today}
	
	\begin{abstract}
		We realize a two-component dipolar Fermi gas with tunable interactions, using erbium atoms. Employing a lattice-protection technique, we selectively prepare deeply degenerate mixtures of the two lowest spin states and perform high-resolution Feshbach spectroscopy in an optical dipole trap. We identify a comparatively broad Feshbach resonance and map the interspin scattering length in its vicinity. The Fermi mixture shows a remarkable collisional stability in the strongly interacting regime, providing a first step towards studies of superfluid pairing, crossing from Cooper pairs to bound molecules, in presence of dipole-dipole interactions.
	\end{abstract}
	
	
	\maketitle

The ability to prepare dipolar quantum gases of magnetic atoms~\cite{Griesmaier2005bec,Lu2011sdb,Lu2012qdd,Aikawa2012bec,Aikawa2014rfd,Naylor2015cdf} has enabled fascinating, yet unexpected, observations, emerging from the long-range and anisotropic character of the dipole-dipole interaction (DDI) among particles.
In bosonic systems with dominant DDI, this includes $d$-wave-patterned collapse~\cite{Lahaye2008dwc}, droplet stabilization~\cite{Kadau2016otr,Chomaz2016qfd,Schmitt2016sbd}, and roton quasi-particles~\cite{chomaz2017oot}.
With fermions, many-body dipolar phenomena have been investigated only in spin-polarized systems. Here, the DDI competes with the Pauli pressure, rendering dipolar effects much more subtle, as e.\,g. their influence on the shape of the Fermi surface~\cite{Aikawa2014oof}.

Magnetic atoms further realize high-spin systems, e.\,g.\,fermionic Er has twenty available spin states in the lowest hyperfine manifold. 
In particular, bosonic dipolar spinor gases have been investigated in remarkable experiments with magnetic Cr atoms~\cite{Hensler2003dri,Pasquiou2011sdo,dePaz2013nqm,Lepoutre2018eoo}, whereas the fermionic counterpart remains rather unexplored in the quantum regime. Scattering experiments with fermionic Dy mixtures slightly above quantum degeneracy, showed a large collisional stability against inelastic dipolar relaxation~\cite{Burdick2015fso}, enabling e.\,g. the production of long-lived spin-orbit-coupled gases via Raman excitations~\cite{Burdick2016lls}.

As yet, the realization of a two-component dipolar Fermi mixture with tunable interactions has remained elusive. Such a system can disclose fascinating phenomena, from anisotropic quantum phases of matter, e.\,g.\, anisotropic Fermi liquids and superfluid pairing~\cite{Baranov2012cmt,Bloch2008mbp}, to dipolar magnetism~\cite{StamperKurn2013sbg}, but also extended Fermi-Hubbard models with off-site interactions~\cite{Dutta2015}. 
Fermionic Er and Dy are very promising candidates for such studies, given their large magnetic moment.
However, the large density of Feshbach resonances (FRs) even in spin-polarized gases~\cite{Frisch2014qci,Baumann2014ool,Maier2015eoc} raises the question of whether stable fermionic quantum mixtures with tunable interactions can be realized with lanthanides.

We here report on a powerful platform to produce a two-component dipolar Fermi gas of pseudo-spin~\nicefrac{1}{2} and demonstrate tunability of the interspin interactions. By using highly magnetic $^{167}$Er atoms and a three-dimensional (3D) optical lattice as a tool for spin preparation, we perform high-resolution Feshbach spectroscopy and unambiguously identify  the spin nature of the different FRs. Among the resonances,  we find a well isolated and comparatively broad interspin FR and precisely measure the interspin scattering length. Our Fermi mixture reveals a remarkable collisional stability in the strongly interacting regime.

Achieving a  deterministic preparation of a spin-\nicefrac{1}{2} mixture and a precise control over the interspin interactions in highly-magnetic lanthanide atoms challenges experimental schemes. Indeed, the enormous density of FRs can cause collisional losses and severe heating, limiting the production and preparation of deeply degenerate mixtures at arbitrary magnetic fields ($B$), where hundreds of FRs might need to be crossed; see e.\,g.\,~\cite{Burdick2016lls}. 
Moreover, state-selective preparation of a spin-\nicefrac{1}{2} system typically requires large $B$ values for which the quadratic Zeeman effect lifts the degeneracy on the Zeeman splitting among consecutive sublevels~\cite{Schmaljohann2004dof,Burdick2015fso}.

For these reasons, we establish a technique for collisional protection during the spin preparation; see Fig.\,1. In a nutshell, the key production steps are: We produce a spin-polarized degenerate Fermi gas (dFg) in an optical dipole trap (ODT) at low $B$ (a1), and load the atoms into the lowest band of a deep 3D optical lattice, which acts as a collisional shield (a2)~\cite{Winkler2007cot,chotia2012lld}. We then sweep to high $B$ for spin preparation and perform radio-frequency (rf) transfer (a3), sweep to the desired $B$ and eventually melt the lattice (a4).

Experimentally, we prepare a spin-polarized dFg of $\Er$ atoms in a crossed-beam ODT~\cite{Aikawa2014rfd,suppmat} (Fig.\,1(a1)). All fermions occupy the lowest Zeeman state $\nine\equiv|F=19/2, m_F=\text{--}19/2\rangle$ of the ground-state manifold. Here, $F$ is the total spin quantum number and $m_F$ its projection along the quantization axis. A homogeneous magnetic field of $B=0.6\,\rm G$ is applied along the vertical $z$ direction to define the quantization axis and to maintain spin polarization. 
The sample typically contains $N=2.4\times 10^4$ atoms at about $T=0.25\, T_{\rm F}$. Note that the ODT is shaped to optimize single-band loading of the optical lattice and yields $E_{\rm F}= k_{\rm B}\times T_{\rm F}=k_{\rm B}\times 170\,{\rm nK}=h\times 3.6\,\rm kHz$~\cite{suppmat}. Here,  $T_{\rm F}$ is the Fermi temperature, $h$ the Planck constant, and $k_{\rm B}$ the Boltzmann constant.

In the next step, we transfer the spin-polarized dFg into a 3D optical lattice (Fig.\,1(a2)), 
{Our lattice has a cuboid geometry with lattice spacings $(d_x,d_y,d_z)\,=\,(266,266,532)\,\mathrm{nm}$ along the three orthogonal directions~\cite{Baier2016ebh,suppmat}.}
In order to pin the atoms in a one-fermion-per-lattice-site configuration (unit filling), we use large lattice depths of about $(s_{x},s_{y},s_{z})\,=\,(20,20,80)$, where $s_i$ with $i\in \{x,y,z\}$ is given in units of the respective recoil energies, $\mathrm{E}_{\mathrm{R}; x,y}=h\times\,4.2\,{\rm kHz}$ and $\mathrm{E}_{\mathrm{R};z}\,=h\times\,1.05\,{\rm kHz}$.
After lattice loading, we obtain a  single-component fermionic band-insulator (BI) of about $2.2\times 10^4$ $\nine$ atoms. 
By melting the lattice and re-loading the fermions into the ODT, we measure $T\lesssim 0.3\,T_{\rm F}$ {with $N=2.1\times 10^4$ ($T_{\mathrm{F}}\approx\unit[160]{nK}$)} and extract a heating rate in the lattice as low as $\Dot{T}=0.03\, T_{\rm F}/\rm s$.

	\begin{figure}
		\includegraphics[width=1\linewidth]{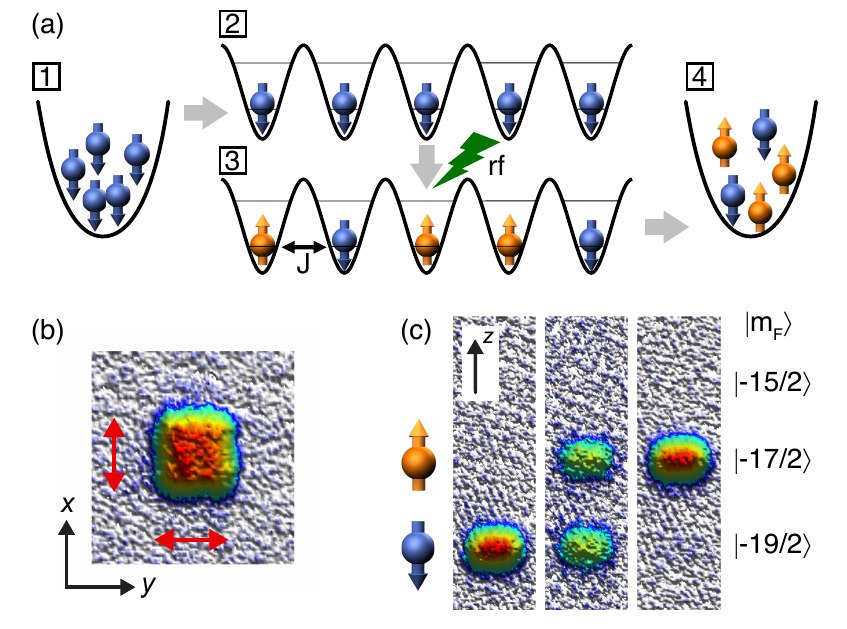}%
		\caption{\label{fig:pol} Spin-\nicefrac{1}{2} dipolar fermions in a 3D optical lattice. (a) Sketch of the four key stages of our preparation scheme; see text. (b) Band population in the horizontal $xy$-plane, obtained by averaging 50 absorption images for a $12\,\rm ms$ time-of-flight (TOF). The red arrows indicate the first Brillouin zone of the lattice. (c) Spin-resolved band-mapping images after $9\,\rm ms$ of TOF in the vertical $z\tilde{x}$ plane, where $\tilde{x}$ accounts for the angle between the imaging beam and the $y$-axis of the lattice, for {population imbalances} $\delta=1$ (left panel), $0.02$ (middle panel), and $-0.94$ (right panel). The images are averages of about $20$ absorption pictures. The spin states are separated along the $z$-direction by a Stern-Gerlach technique.}
	\end{figure}

Our system is well described by a single-band extended Fermi-Hubbard model~\cite{Dutta2015} with residual tunneling rates of $J_{x,y}=h\times 10.5\,\rm Hz$ and $J_{z}=h\times 0.001\,\rm Hz$, and nearest-neighbor interactions in the order of  $h\times 50\,\rm Hz$~\cite{Baier2016ebh}. We confirm the single-band population by performing standard band-mapping measurements~\cite{Kastberg1995aco}.
In the horizontal ($xy$) plane, we do not resolve higher-band occupation; see  Fig.\,1(b)~\cite{suppmat}.
Along the $z$-axis, we detect a residual $< 5\,\%$ population in the first excited band, resulting from the fact that $E_{\rm F}>E_{\mathrm{R},z}$~\cite{Will2011thesis}. Because of the Pauli exclusion principle, doubly-occupied sites (doublons) in a single band are strictly forbidden for identical particles ($\nine$).

In the BI regime, the lattice is expected to provide a strong collisional protection to the particles. 
As a first application, we use the lattice-protection technique to realize a spinor Fermi gas with pseudo-spin~\nicefrac{1}{2} ($\nine$--$\seven$), with $\seven\equiv|F=19/2, m_F=\text{--}17/2\rangle$ (Fig.\,1(a3)). 
Experimentally, we start with a $\nine$ BI at $B=0.6\,\rm G$ and then ramp $B$ in $40\,\rm ms$ to a value of about $40\,\rm G$, for which the quadratic Zeeman effect in $^{167}\rm Er$ is large enough to lift the degenerate coupling of the individual spin levels~\cite{suppmat}. After letting the field stabilize for $120\,\rm ms$, we use a standard rf-sweep technique to transfer part of the atoms into the $\seven$ state. By tuning the rf-power, we can precisely control the population imbalance, $\delta = (N_{\down}-N_{\up})/N$, in the mixture, with $N_{\down}$ ($N_{\up}$) the number of atoms in $\nine$ ($\seven$). Figure 1(c) shows exemplary spin-resolved absorption images of $\nine$--$\seven$ mixtures for various $\delta$ after $B$ is swept back to low values. {We typically record $N=N_{\down}+N_{\up}=1.8\times10^4$  and $T\approx 50\,$nK after melting the lattice down (Fig.\,1(a4)).}
For comparison, similar measurements in absence of the lattice clearly show a much lower atom number of $N=0.6\times10^4$, proving the strength of our lattice-protection scheme to circumvent losses when cruising through the ultra-dense Feshbach spectrum~\cite{Burdick2016lls,Frisch2014qci}.

Figure 2 shows the high collisional stability of the lattice-confined spin mixture. 
In particular, we probe $N_{\down,\up}$ as a function of the holding time in the lattice; see Fig.\,2(a). From an exponential fit to the data, we extract long lifetimes of $\tau_{\down}=\,31(3)\,\rm s$ and $\tau_{\up}=\,12.2(7)\,\rm s$. The measurements are carried out at $B=3.99\,\rm G$, where no FRs occur~\cite{suppmat}. 
{Interestingly, within our error bars, we find no dependence of the lifetime of each spin state on the population in the other state; they remain long regardless of $\delta$; see Fig.\,2(b).}

We note that, although very long for our purpose, we always record shorter lifetimes for a $\seven$ BI with respect to the ones measured for a $\nine$ BI. Differently from the $\nine$ case, two-body relaxation processes for $\seven$ are allowed. At our magnetic fields, this process converts Zeeman energy into a large enough kinetic energy to let the atoms escape from the lattice ~\cite{Hensler2003dri,dePaz2013rdo}, and requires the particles to collide at short distance (onsite)~\cite{Pasquiou2010dir,Burdick2015fso}. In the spin-polarized cases (e.\,g.\, $\delta=-1$; $\seven$), double occupancy necessarily involves population in higher bands since the Pauli exclusion principle forbids doublons in the lowest band. In our system, a continuous transfer of a small fraction of atoms into higher bands might be driven by intensity and frequency noise of the lattice beams~\cite{chotia2012lld}. In the case of $\seven$ this would lead to subsequent fast relaxation and justify the observed difference in the lifetimes. 

	\begin{figure}
    	\includegraphics[width=1\linewidth]{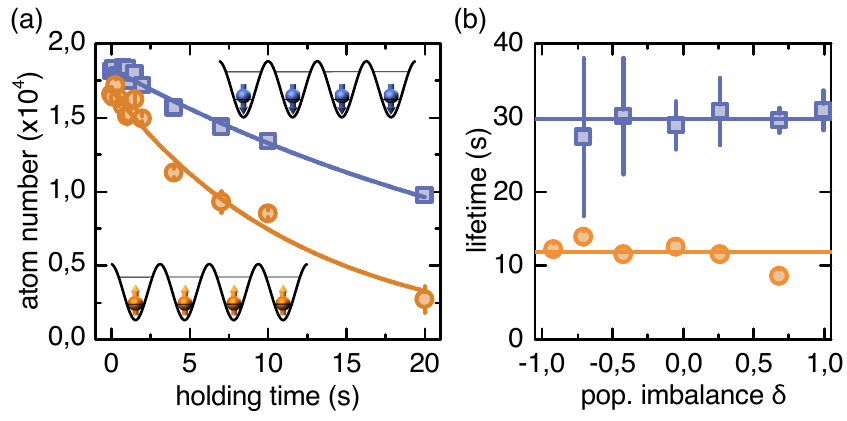}%
    	\caption{\label{fig:mixture} Spin mixture of dipolar $^{167}$Er in a 3D lattice. (a) Lifetime measurements for spin-polarized samples of $\nine$ (squares) with $\delta=1$ and of $\seven$ (circles) with $\delta=-0.92$ at $B=3.99\,\rm G$ and their respective exponential decay (solid lines). (b) Lifetimes as a function of $\delta$. 
    	Constant fits extract mean lifetimes across $\delta$ of $\bar{\tau}_{\down}=29.9(3)\,\rm s\,$ and $\,\bar{\tau}_{\up}=11.8(7)\,\rm s$. 
    	All error bars indicate the statistical uncertainty.}
	\end{figure}

With our spin-preparation method, we are now able to conduct high-precision Feshbach spectroscopy in an ODT (Fig.\,1(a4)) in search of interspin loss features.
For this, we first prepare the spin-\nicefrac{1}{2} mixture in a deep lattice at the desired $B$ value. We then transfer the mixture back into the ODT, hold the atoms for $500\,\rm ms$, and finally measure the spin populations.
Figure 3 exemplifies the high-precision Feshbach spectroscopy for three values of $\delta$ within a narrow magnetic field range from $B=550\,\rm mG$ to $750\,\rm mG$ with a resolution of $1\,\rm mG$. A lower-resolution and larger-range scan is shown in Ref.\,~\cite{suppmat}. 
   
  	\begin{figure*}[ht]
		\includegraphics[width=1\linewidth]{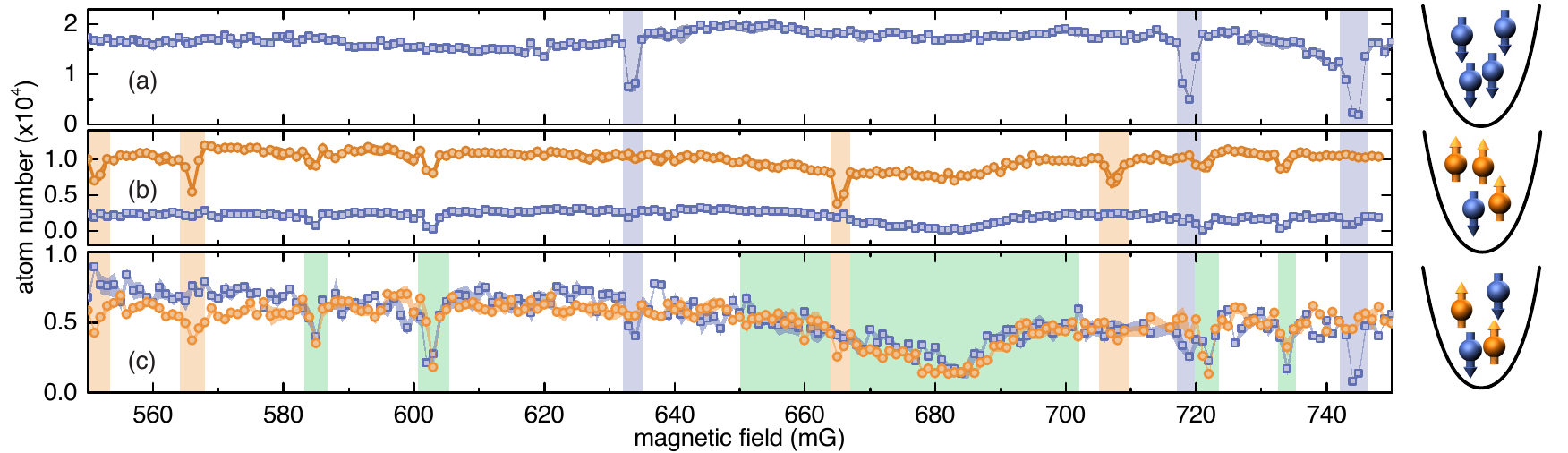}
	    \caption{\label{fig:feshbach} 
		High-resolution Feshbach spectroscopy for three different population imbalances in an ODT (cartoons): atoms in $\nine$ (squares) and $\seven$ (circles) for
	    $\delta=1$ (a), $-0.6$ (b), and $0$ (c) as a function of $B$.
	    The determined width and spin nature of the FRs are indicated by the blue ($\nine$), orange ($\seven$), and green ($\nine$--$\seven$) shaded regions. {Each data point is the mean of two to four repetitions.}}
	\end{figure*}
    
As expected~\cite{Frisch2014qci,Baumann2014ool}, the atom-number trace as a function of $B$ shows a high density of resonant loss features on top of a constant background. By controlling $\delta$, we are able to distinguish the spin nature of each of the observed FRs. In the excerpt shown in Fig.\,3, we identify three narrow homo-spin FRs in a pure $\nine$ sample (a) and four in a quasi-pure $\seven$ sample (b).  In the spin-polarized cases, all FRs exhibit widths of the order of our magnetic field stability of $\approx 1\,\rm mG$. Thanks to our lattice-preparation technique, the shape and the width of the FRs are not affected  by the magnetic field ramps, namely we do not observe neither broadening nor fictitious asymmetry in the loss peaks. 
For the $50\%$-$50\%$ spin mixture ($\delta=0$), we observe five additional interspin FRs (Fig.\,3(c)), where atoms in the two spin states are simultaneously lost. Because of the complicated scattering behavior of Er, standard coupled-channel methods to assign the leading partial-wave character of the FRs are currently not available~\cite{Frisch2015upm}. However, the width of the FRs can give indications on the strength of the coupling between open and closed channels~\cite{Chin2010fri}. 

Among the observed interspin FRs, the one at about $0.68\,\rm G$ stands out from the forest of narrow FRs. This FR is almost two orders of magnitude broader, making it a promising candidate for Fermi-gas experiments in the strongly interacting regime. We further investigate this FR 
by performing modulation spectroscopy on the lattice-confined spin-\nicefrac{1}{2} mixture (Fig.\,1(a3)) to extract the interspin onsite interaction energy, $U_{\down\up}=U_{\mathrm{c}}+U_{\mathrm{dd}}$, given by the sum of the interspin contact interaction, $U_{\mathrm{c}}$, and the DDI, $U_{\mathrm{dd}}$~\cite{Baier2016ebh}.
{Thanks to the precise knowledge of $U_{\mathrm{dd}}$ and to its angle dependence, we are able to directly extract the interspin scattering length, $a_{\down\up}\propto U_{\mathrm{c}}=U_{\down\up}-U_{\mathrm{dd}}$, both in amplitude and in sign; for details see Ref.\,\cite{suppmat}.} 

{Figure 4(a) summarizes our results, showing the tunability of $a_{\down\up}$ from positive to negative values across the interspin FR. As a first estimate of the $B$-to-$a_{\down\up}$ conversion, we use the simple single-channel formula, leading to $a_{\down\up}(B)=a_{\rm bg}\left(1-\frac{\Delta}{B-B_0}-\frac{\Delta^{\prime}}{B-B_0^{\prime}}\right)$~\cite{Chin2010fri}. 
From the fit to the data, we extract the background scattering length $a_{\rm bg}=91(8)\,a_0$, the position of the comparatively broad FR $B_0=687(1)\,\rm mG$, and its width $\Delta=58(6)\,\rm mG$. 
Note that our fitting function also accounts for a nearby interspin FR at $B_0^{\prime}=480\,\rm mG$ ~(out of range of Fig.\,3 and 4) of width $\Delta^{\prime}=29(4)\,\rm mG$, whereas narrower interspin FRs are neglected.
Based on the extracted values, we can estimate {an order of magnitude for the effective range of the FR}, $R^*=\hbar^2/(m_{\mathrm{Er}}\Delta a_{\rm bg}\delta\mu$) ~\cite{Chin2010fri}. Here, $m_{\mathrm{Er}}$ is the mass of $^{167}$Er. The differential magnetic moment between the open and closed channel, $\delta\mu$, is not known for the considered FR. 
However, taking $\delta\mu= 3\,\mu_{\mathrm{B}}$, which is the typical value measured on bosonic Er$_2$~\cite{Frisch2015upm}, we estimate $R^*$ on the order of $1000 a_0$. With this order of magnitude, our typical two-component Fermi gases verify $1/k_{F}R^*\gtrsim 1$ with $k_{\rm F}$ being the Fermi wave vector~\cite{suppmat}. This identifies
the intermediate strength of the FR~\cite{Gurarie2007rpf}, for which the gas is expected to remain strongly interacting at unitarity~\cite{Ho2012art, Hazlett2012roa}.}

	    \begin{figure}
	\includegraphics[width=1\linewidth]{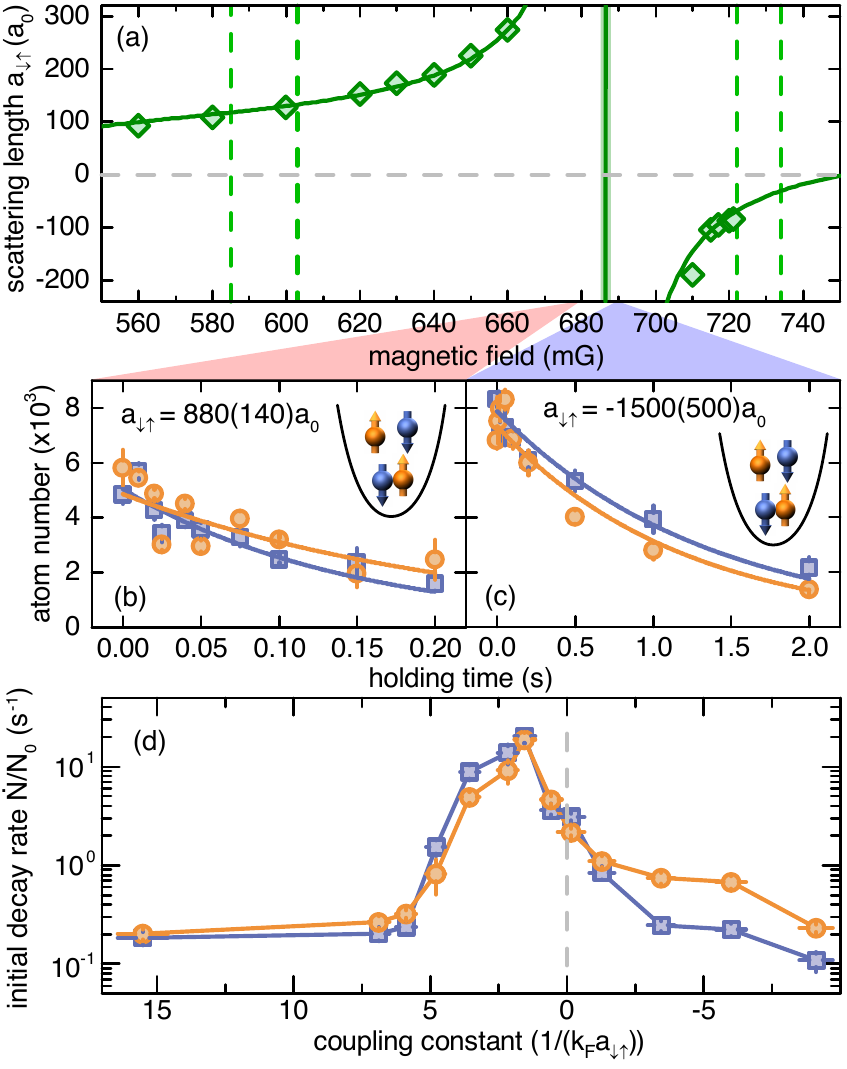}%
	\caption{\label{fig:scattprop} Interspin scattering length and collisional behavior of the strongly interacting Fermi mixture (a) $a_{\down\up}$ extracted from modulation spectroscopy in the lattice. Error bars are smaller than the size of the data points.
	Vertical dashed lines indicate the position of narrow interspin FRs as identified in Fig.\,3(c).	The solid line is a fit to the data with the fit statistical uncertainty indicated as shaded region.
	(b-c) Lifetimes of a spin mixture of $\nine$ (squares) and $\seven$ (circles) with $\delta=0$ in an ODT for large positive $a_{\down\up}$ at $680\,\rm mG$ (b) and for large negative  
	$a_{\down\up}$ at $ 690\,\rm mG$. (d) Initial decay rate $\dot{N}/N_0$ of the normalized atom numbers as a function of $1/(k_{\rm F}a_{\down\up})$ in the vicinity of the FR. {Error bars on the coupling constant are deduced via standard error propagation.}}
    \end{figure}

For strongly interacting alkali Fermi gases, the large collisional stability in  two-component mixtures has been essential for observing the crossover from a superfluid of delocalized pairs along the Bardeen-Cooper-Schrieffer (BCS) mechanism to a Bose--Einstein condensate (BEC) of bound molecules~\cite{Varenna2007ucf}. As a direct consequence of the Pauli principle, three-body recombination occurs primarily on the repulsive (BEC) side of broad $s$-wave FRs, where a weakly bound molecular level exists~\cite{Petrov2003tbp}, whereas, on the attractive (BCS) side, large scattering lengths coexist with a remarkable collisional stability ~\cite{Dieckmann2002doa,Bourdel2003mot,Regal2004lom,Jochim2004thesis}. Such an asymmetry in the scattering behavior is identified as an essential attribute of BEC-BCS physics. 


We investigate this aspect in a second set of experiments. We prepare an equally populated spin mixture ($\delta=0$) in an ODT (Fig.\,1(a4)) and probe the time evolution of the spin population as a function of the holding time in the trap for various $B$ across the FR. Exemplary decay curves are shown in Fig.\,4(b-c). On the BEC side, at $a_{\down\up}=880(140)\,\rm a_0$, we observe a fast decay of both $\seven$ and $\nine$ atoms (Fig.\, 4(b)). A simple exponential fit to the data gives lifetimes of $\tau_{1/e}\approx 150\,\rm ms$. 
In contrast, on the BCS side at $a_{\down\up}=-1500(500)\,\rm a_0$ (Fig.\, 4(d)), the spin mixture shows a large collisional stability with lifetimes exceeding $\tau_{1/e}= 1200\,\rm ms$ (Fig.\, 4(c)). 
	
To get deeper insights, we systematically study the initial decay rate, $\dot{N}/N_0$, as a function of $B$. We determine the rates by using a linear fit to the data for the initial time evolution. Figure 4(d) summarizes our results, plotted in terms of the dimensionless coupling constant $1/(k_{\rm F}a_{\down\up})$. {We observe an asymmetry of the loss rate curve, indicating that the Fermi mixture is remarkably stable in the unitary and strongly attractive regime. We note that both the qualitative shape and the quantitative values of the loss rates in $^{167}$Er show strong similarities to the ones measured in $^{40}{\rm K}$~\cite{Regal2004lom}.}
	
{The existence of a comparatively broad interspin FR, and our demonstration of the interaction tuning across this resonance make fermionic Er gases a promising system for accessing BEC-BCS crossover physics within a distinct scattering scenario. Indeed, our mixture adds both the DDI and an intermediate effective range in the short-range scattering compared to the alkali cases~\cite{Varenna2007ucf, Hazlett2012roa}, paving the way for studying exotic Cooper pairs and molecular BECs~\cite{Baranov2012cmt,Schwenk2005rfg,Gurarie2007rpf} and calling for new theory developments~\cite{Ming2018oqm}.}

\begin{acknowledgments}
We thank R.\,Grimm, M.\,Greiner, and A.\,M.\,Rey for fruitful discussions.
We acknowledge financial support through an ERC Consolidator Grant (RARE, no.\,681432), FET Proactive project (RySQ, no.\,640378) of the EU H2020, and a DFG/FWF (FOR 2247). LC is supported within a Marie Curie Project (DipPhase, no.\,706809) of the EU H2020. 
\end{acknowledgments}

$^\dagger$ Current address: Physikalisches Institut, Universit\"at Heidelberg, Im Neuenheimer Feld 226, 69120 Heidelberg, Germany

* Correspondence and requests for materials
should be addressed to F.F.~(email: francesca.ferlaino@uibk.ac.at).

\providecommand{\noopsort}[1]{}\providecommand{\singleletter}[1]{#1}%
%


\appendix
\renewcommand\thefigure{\thesection S\arabic{figure}}   
\setcounter{figure}{0}   
\section{Supplemental Material}
\subsection{Spin-polarized degenerate Fermi gases}
Our experimental protocol for the preparation of deeply degenerate Fermi gases (dFgs) of $\Er$ follows the one described in ref.\,~\cite{Aikawa2014rfd}. The experiment starts with a narrow-line magneto-optical trap operated at $583\,\rm nm$ to prepare spin-polarized $\Er$ atoms with $N= 1.2\times 10^7$ atoms  {and $T\approx 10\mu K$} in the lowest hyperfine sublevel $|F=19/2, m_F=\text{--}19/2\rangle$, where $F$ is the total angular momentum quantum number and $m_F$ is its projection along the quantization axis. 
The atoms are then transferred to a horizontal optical dipole trap (ODT) formed by a laser beam at $1064\,\rm nm$. The aspect ratio $\mathrm{AR}=w_{\perp}/w_z$ between the horizontal, $w_{\perp}$, and vertical, $w_z$, waists of this beam can be tuned from $1.6$ to $15$ via a time-averaging potential technique~\cite{Chomaz2016qfd}, which allows to reach a good spatial mode overlap between the atomic cloud and the ODT. 
Subsequentially the atomic cloud is compressed by reducing the $\rm AR$ and transferred to a tight ODT created by a laser beam at $1570\,\rm nm$ with a waist of about $15\,\rm \mu m$, and counterpropagating to the $1064\,\rm nm$-beam such that their focii overlap.
At this stage we typically have $1\times 10^6$ atoms. 
During the evaporation procedure the atoms are further confined by an additionnal ODT at $1570\,\rm nm$, formed by a beam propagating vertically with a waist of about $32\,\rm \mu m$. The crossed ODT at $1570\,\rm nm$ is later denoted $\HC$.

Following our previous work of ref.~\cite{Aikawa2014rfd}, we perform evaporative cooling based on elastic dipolar scattering among identical fermions. Such a cooling scheme has been proven to be very efficient to produce samples in the deeply quantum degenerate regime~\cite{Aikawa2014rfd,Burdick2016lls}. 
At the end of the evaporation, the trap frequencies in $\HC$ are $(\nu_{\perp},\nu_{\parallel},\nu_{z})=(286(3), 85(1), 255(3))\,\rm Hz$  with $\parallel$ ($\perp$) corresponding to the axis along (perpendicular to) the horizontal ODT beam and $z$ indicating the axis of gravity. We typically obtain spin-polarized dFgs with up to $N=6\times 10^4$ atoms and temperatures of $T\leq 0.15\, T_{\rm F}$, with $T_{\rm F}$ being the Fermi temperature corresponding to the Fermi energy $E_F=k_{\rm B}T_{\rm F}=h \bar{\nu}(6N)^{1/3}$, where $h$ is the Planck constant, $\bar{\nu}=\left(\nu_{\perp}\nu_{\parallel}\nu_{z}\right)^{1/3}$ the geometric mean of the trap frequencies and $N$ the atom number. At this stage the Fermi energy is $E_F=k_{\rm B}\times 630\,{\rm nK}=h\times 13\,\rm kHz$. During the whole evaporation, the magnetic field has a value of $B=0.6\,\rm G$ and is oriented along $z$, which sets the quantization axis of the atomic dipoles. Here and in the following, $N$ and $T/T_{\rm F}$ are estimated from polylogarithmic fits to the absorption images of the dFGs after 12\,ms of time-of-flight (ToF) using the horizontal imaging setup.

\subsection{Preparation for lattice loading}

In deeply dFgs, the atoms fill the Fermi sea up to the Fermi energy, $E_F$. Hence, the number of populated bands, when the atoms are loaded to an optical lattice, crucially depends on the initial value of $E_F$. In first approximation, $E_F$ can be compared to the lattice recoil energy $E_{\rm rec}=h^2/(2m_{\rm Er}\lambda^2)$, with $m_{\rm Er}$ being the mass of $\Er$ and $\lambda$ the lattice wavelength. In particular, during the initial increase of the lattice potential higher bands become populated if $E_F> E_{\rm rec}$~\cite{Will2011thesis}.

To minimize the occupation of higher bands due to the loading procedure we reduce the Fermi energy of our sample. To this aim, we transfer the atoms back to a crossed ODT operated at  $1064\,\rm nm$ ($\mopa$), within $510\,\rm ms$. Here, the dynamically adjustable $\rm AR$ of the horizontal beam allows a convenient control on $\bar{\nu}$ (see section above). We optimize the ODT parameters by lowering $\bar{\nu}$ and $N$ while keeping a low temperature of the sample. The best conditions for subsequent lattice loading are reached for 
$(\nu_{\perp},\nu_{\parallel},\nu_{z})=(63(1),36(2),137(1))\,\rm Hz$ and $N=2.4\times 10^4$ atoms with $T\leq 0.3\, T_{\rm F}$, corresponding to a Fermi energy $E_{\rm F}= k_{\rm B}\times 170\,{\rm nK}=h\times 3.6\,\rm kHz$.
We note that for lower $\nu_{z}$ atoms get lost due to gravity.

\subsection{Three-dimensional optical lattice and its loading}
 The  three-dimensional (3D) optical lattice in our experiment is created by two retro-reflected $\unit[532]{nm}$ laser beams along the $x$- and $y$-axis and one retro-reflected $\unit[1064]{nm}$ vertical laser beam along the $z$ axis; see Fig.\,S1. The lattice spacings are $d_{x,y}\,=\,266\,{\rm nm}$ along the horizontal $xy$-plane and $d_{z}\,=\,532\,{\rm nm}$  along the vertical $z$-axis~\cite{Baier2016ebh}. With the available power, we reach maximum lattice depths of $(s_{x},s_{y},s_{z})\,=\,(25,25,120)$, where $s_i$ with $i\in \{x,y,z\}$ is given in units of the respective recoil energies, $\mathrm{E}_{\mathrm{R}; x,y}=h\times\,4.2\,{\rm kHz}$ and $\mathrm{E}_{\mathrm{R};z}\,=h\times\,1.05\,{\rm kHz}$. 
 Typical lattice depths used in the experiment are $(s_{x},s_{y},s_{z})\,=\,(20,20,80)$ corresponding to band gaps of $h\times\,32.8\,\rm kHz$ along $x$ and $y$ and $h\times\,17.7\,\rm kHz$ along $z$. 
 
 After preparation and transfer to the $\mopa$, we adiabatically load the spin-polarized dFg into the 3D lattice by increasing the lattice-beam intensities exponentially in $150\,\rm ms$ to the final values. Subsequently, the ODT beams are switched off in $\unit[10]{ms}$ and we additionally hold the atoms for $500\,\rm ms$ before applying our spin preparation scheme. This holdtime enables to remove most of the residual atoms that have been pushed to higher bands of the optical lattice by the Fermi pressure, through their natural faster decay. We note that when the atoms are loaded directly from $\HC$ we find up to $25\,\%$ of population in higher bands, which in this case get strongly depopulated within $500\,\rm ms$. Despite our most careful loading procedure and our holdtime, we measure that up to $5\,\%$ of the atoms can still populate the higher band of the vertical lattice; see main manuscript (Note that the higher bands along $z$ are the most tightly trapped within our lattice geometry).

 \subsection{Higher-band populations}
 
To access the band population we perform band-mapping measurements. Here, we decrease all lattice potential to zero within $1\,\rm ms$, thus mapping the quasi-momentum of the band to real momentum. We then perform  ToF absorption imaging, which thus probes the population of the different bands in directions transverse to the imaging axis.
We note that, the edges of the lowest band can be smeared out because of the finite width of the in-situ cloud and due to an imperfect adiabatic switch-off of the lattice potentials~\cite{Koehl2005fai}, limiting the accuracy of our determination. {In our setup, we obtain our best estimate of the remaining higher band populations by comparing the absorption images to the expected profiles computed from the first Brillouin zone.
In the $z$-direction, we observe a very small population in higher bands, which we quantitatively estimate using the horizontal imaging setup. Here, we additionally take advantage of the observed structure of the higher band population, which systematically appears below the lowest band (along the gravity axis). 
This might be due to a combined effect of residual magnetic gradient and gravity. We use this behavior to our advantage and extract the population of the higher band in $z$ from the top-bottom asymmetry of the band-mapping images.
In the $xy$-directions, the estimate of the higher band population is more subtle in particular because of the non-orthogonal configuration between the imaging and horizontal lattice axes. To the best of our detection sensitivity, we do not observe any population in higher-bands along these axes. Based on the higher recoil energies in these directions, we physically expect a lower initial population of those bands than along $z$, as discussed above. In addition, because of the lower lattice depths, we expect a faster loss of their population.}

\begin{figure}
	\includegraphics[width=1\linewidth]{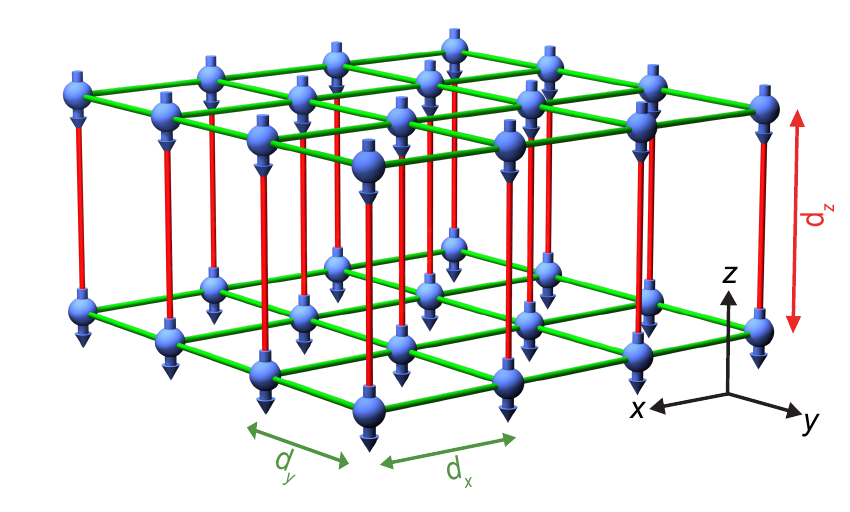}%
	\caption{\label{fig:latt}
	Sketch of our lattice geometry. The coordinate system $\{x,y,z\}$ and the lattice constants $d_x$, $d_y$, and $d_z$ are indicated.
	}
\end{figure}

\subsection{Zeeman energy for fermionic Er}
Fermionic Er exhibits a hyperfine structure resulting from the coupling of its nuclear spin $\bf I$, whose quantum number is $I=7/2$, with the total electronic angular momentum $\bf J$, which in the ground state of Er has for quantum number $J=6$. The total angular momentum reads $\bf F=J+I$. In the lowest hyperfine manifold ($F=19/2$) there are $2F+1=20$ sublevels which can be differentiated by the eigenvalues of the projection of $\bf F$ on the quantization axis, corresponding to the quantum number  $m_F$. Because, in our experiment, an external magnetic field $\bf B$ is always applied, the degeneracy of the sublevels is lifted by the interaction of $\bf B$ with the different angular momenta.  In our description, the quantization axis is chosen to be parallel to $\bf B$, and the $m_F$ sublevels are then denoted magnetic states. In the low $B$ limit, the magnetic states are simply shifted in energy along $E_{m_F}=m_Fg_F\mu_B B$, corresponding to a state dependent magnetic moment, $\mu = m_Fg_F\mu_B$. In Erbium, the absolute ground state has a magnetic moment of $\mu =-6.982804 \mu_B$, giving the Land\'e factor $g_F=0.735032$~\cite{Conwaylle1963}. Here $\mu_B$ is the Bohr magneton.

In a more general way, the atomic energy levels in a uniform $B$-field can be calculated via an exact diagonalization of the atomic Hamiltonian~\cite{Smith1965ths}. In Fig.\,S2 we plot the energy levels of the lowest hyperfine manifold as a function of the magnetic field computed in such a way. The dominant trend evidences the linear dependence discussed above in the low $B$ regime. However, at large enough $B$, deviations from this simple picture appear due to the Paschen-Back effect, as $\bf J$ and $\bf I$ start to decouple. In a perturbative description, this can be accounted via a quadratic correction to the Zeeman energies which writes $E_{m_F} - q_{\rm{li}}m_F= q_{\rm{qu}}(m_F^2-F^2)$ with $q_{\rm{li}} = g_F\mu_B B$  and $q_{\rm{qu}}\propto B^2$. The deviation from the linear Zeeman energy becomes evident when considering the differential splitting $\Delta E_{\rm Z}(m_F)=(E_{m_F}-E_{{m_F+1}})-(E_{{m_F+1}}-E_{{m_F+2}})$, as the linear Zeeman effect $q_{\rm{li}}m_F$ cancels out; see inset of Fig.\,S2. In the folowing we define $\Delta E_{\rm Z}=\Delta E_{\rm Z}(m_F=-19/2)$, which is the most relevant quantity for the current study, as restricted to $m_F=-19/2$ and $-17/2$.  

\begin{figure}
	\includegraphics[width=1\linewidth]{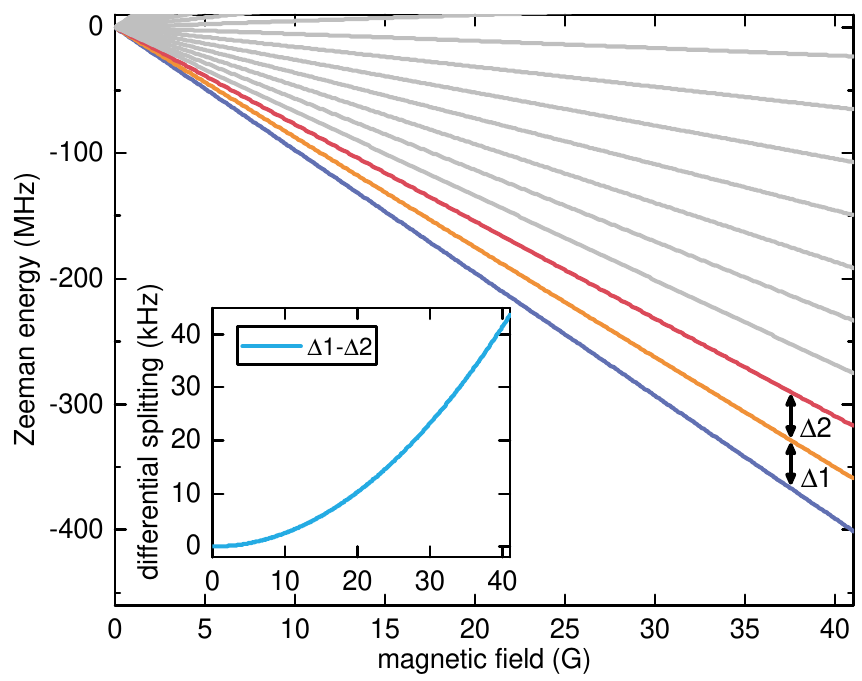}%
	\caption{\label{fig:Zeeman}
	Zeeman energy for the magnetic substates in the $|F=19/2\rangle$ hyperfine manifold. For this work the energy splitting of the lowest three spin states $|\text{--}19/2\rangle\equiv\nine$ (blue line), $|\text{--}17/2\rangle\equiv\seven$ (orange line) and $\five$ (red line) is of most relevance. Higher spin states are visualized by grey lines. The linear Zeeman effect dominates the energy evolution such that $\Delta1\approx \Delta2 \approx q_{\rm{li}}$ while the quadratic Zeeman effect is evident in the differential splitting $\Delta1-\Delta2$ (inset).
	}
\end{figure}

\subsection{Preparation of a spin mixture in the lattice}
To achieve a deterministic spin preparation of the two lowest spin states we typically use a large enough magnetic field of $B=40.51\,\rm G$ for which the differential Zeeman splitting $\Delta E_{\rm Z} = 42.6\,\rm kHz$ is larger than the fluctuations of the Zeeman energies coming from the magnetic field noise, corresponding to $\approx 20\,\rm kHz$ at this field. In this way, the spin-spin coupling induced by a small amplitude modulation of the magnetic field in the radio-frequency (rf) domain can be restricted to the subspace formed by the lowest two spin states $m_F=-19/2$ and $-17/2$.  
To couple the two hyperfine sub-states we apply a rf-sweep by chirping the rf-frequency continuously from a value of $(\nu^{\star}+30\,\rm kHz)$ to $(\nu^{\star}-30\,\rm kHz)$ within about $10\,\rm ms$, where $h\nu^{\star}$ matches the energy difference $E_{-19/2}-E_{{-17/2}}$ ($\Delta 1$ in Fig.\,S2).
We can prepare a well-reproducible mixture of $\nine$ and $\seven$ without populating the next higher spin state. 
The population imbalance $\delta$ between the two spin states can be freely controlled by varying the power of the rf signal. In particular, also almost all the atoms can be transferred to $\seven$ reaching up to $\delta=-0.94$ (see Fig.\,1(c) of the main manuscript). 
We note that, while our preparation technique in the lattice initially leads to a coherent superposition of the two spin states, additional measurements suggest a fast decoherence, leading to a projection of pure states on the individual lattice sites for the experimental relevant time scales. In particular, we observe that coherently driven Rabi oscillations between the two spin states quickly damp within a few $\rm ms$.

To image the spin mixture, we perform spin-resolved TOF absorption imaging using a Stern-Gerlach technique with a $1\textrm{-ms}$ pulsed magnetic field gradient at the beginning of the TOF. After an additional $7.2\,\rm ms$ of TOF, the populations of each spin state are spatially separated and we measure them by using the horizontal imaging setup (Fig.\,1(c)). $N_{\down.\up}$ are then counted by integrating the measured density distribution over well-defined regions of interest.

\subsection{Lifetime of the spin mixture in a deep lattice}
To conduct a clean measurement of the collisional properties of a spin mixture in the deep optical lattice it is important to fulfill the following requirements:
(i) The spin mixture is in an insulating regime where the formation of doublons is suppressed via sufficiently large ratios of the onsite interspin interaction energy $U_{\up\down}$ to the tunneling rate $J$. This requires not only to use a deep lattice potential but also to sit away from any Feshbach resonance (FR) so that $U_{\up\down}$ is not resonantly modified and has a value close to its background one. 
(ii) The Zeeman energies are large and do not have an equidistant spacing so that both magnetization changing and magnetization conserving spin-exchange processes induced by the DDI are energetically supressed~\cite{dePaz2013nqm,Lepoutre2018eoo,Yan2013ood,Baier2017prep}. 

Due to the high density of FRs (see next section and Ref.\,\cite{Frisch2014qci}), the requirement (i) is not so straightforwardly achieved in our fermionic erbium mixture. In this system, 
 to find a magnetic field value for which we sit stably away from any FR is eased by a low technical magnetic-field noise. Because of the different sets of coils used in our experiment, the magnetic field noise is found to be $\approx 1\,\rm mG$ up to $B=5\,\rm G$ while it increases to $\approx 20\,\rm mG$ when we go to higher $B$ values. Hence, working at $B\leq 5\,\rm G$ turns out to be more favorable in our setup.
On the other hand, the requirement (ii) is matched for a sufficiently large magnetic field, where the quadratic Zeeman effect (see Fig.\,S2) is strong enough to not be canceled by quadratic light shifts~\cite{Becher2018apo}.
The best conditions for meeting the requirements (i) and (ii) are found at $B=3.99\,\rm G$, which is used for the lifetime measurements of Fig.\,2. At this field, we have measured the onsite interaction energy using a similar technique as for Fig.\,4(a) of the main text; see also below. The extracted value of $U_{\up\down}=h\times2.43(2)\,\rm kHz$ exceeds by far the relevant tunneling rates $J_{x,y}=h\times 10.5\,\rm Hz$ so that it lies deeply in the insulating regime. In addition, we note that no spin-changing dynamics are observed from the measured spin population.

{We finally note that, for technical reasons, our setup allows holding in the lattice up to 20 seconds. For longer times, thermal effects in the high-power fiber start to prevent us from properly stabilizing the output power for our lattice. We avoid observation in this regime where the lattice potential would not be properly controlled (which may also lead to additional losses), as the allowed observation time is already very long for our purposes. This restriction however limits the precision of our lifetime measurements for the long lifetimes observed in our setup and in particular for the longest-lived $\nine$ state.}

\subsection{State-resolved Feshbach spectroscopy}
To identify the magnetic field regions where promising interspin FRs occur, we first perform a rough Feshbach scan in the $0-2\,\rm G$ region for different population imbalances $\delta$ (Fig.\,S3).  For this set of data we do not use our lattice-protection technique. Instead, the spin preparation, the magnetic-field ramps, and the Feshbach spectroscopy are directly performed in the ODT. As expected, without the lattice, the loss features present broadening and asymmetric shapes due to the mere magnetic-field sweeps (e.\,g.\,via losses occuring during the sweeps). Yet, the most prominent features of the scattering physics can be identified. 

Using this technique, we perform three sets of measurements, varying the composition of the mixture $\delta$. In a first set, we perform a Feshbach scan in a spin polarized gas in $\HC$ (Fig.\,S3, upper panel). We jump to the final magnetic field and hold for $t_{\rm hold}=70\,\rm ms$ before TOF imaging. The trap frequencies are $(\nu_{\perp},\nu_{\parallel},\nu_{z})=(324(1), 147(5), 259(4))\,\rm Hz$. The system has an initial temperature of $T=0.18(1)\,T_{\rm F}$. Similar to Ref.\,\cite{Frisch2014qci}, we observe a high density of loss features, which correspond to single-component ($\nine$) FRs of high partial-wave character.
In a second set of measurements, we repeat the magnetic-field scan in an almost pure $\seven$ sample (Fig.\,S3, middle panel). Here, we use a resonant rf-pulse at $0.99\,\rm G$ to prepare a mixture with mainly $\seven$ atoms. Then we jump on a purely {$\nine$} homo-spin FR located at $1.034\,\rm G$ to remove remaining $\nine$ atoms. The measurement is performed in the more shallow $\mopa$ to prevent too strong interspecies losses and $t_{\rm hold}=500\,\rm ms$. For this trap, the trap frequencies are $(\nu_{\perp},\nu_{\parallel},\nu_{z})=(39(1), 37(1), 145(3))\,\rm Hz$ and the initial temperature is $T=0.35(1)\,T_{\rm F}$. We find new FRs, which mainly correspond to single-component $\seven$ FRs. 
In a third set of measurements, we observe the loss features for a spin mixture prepared at $0.58\,\rm G$ in the same trap as for the pure $\nine$ measurement with $t_{\rm hold}=50\,\rm ms$ (Fig.\,S3, lower panel). Here, the initial temperature is slightly increased to $T=0.24(1)\,T_{\rm F}$  due to the spin mixing. The individual homo-spin FRs are still visible while we also find new interspin $\nine$--$\seven$ FRs.

\begin{figure*}
	\includegraphics[width=1\linewidth]{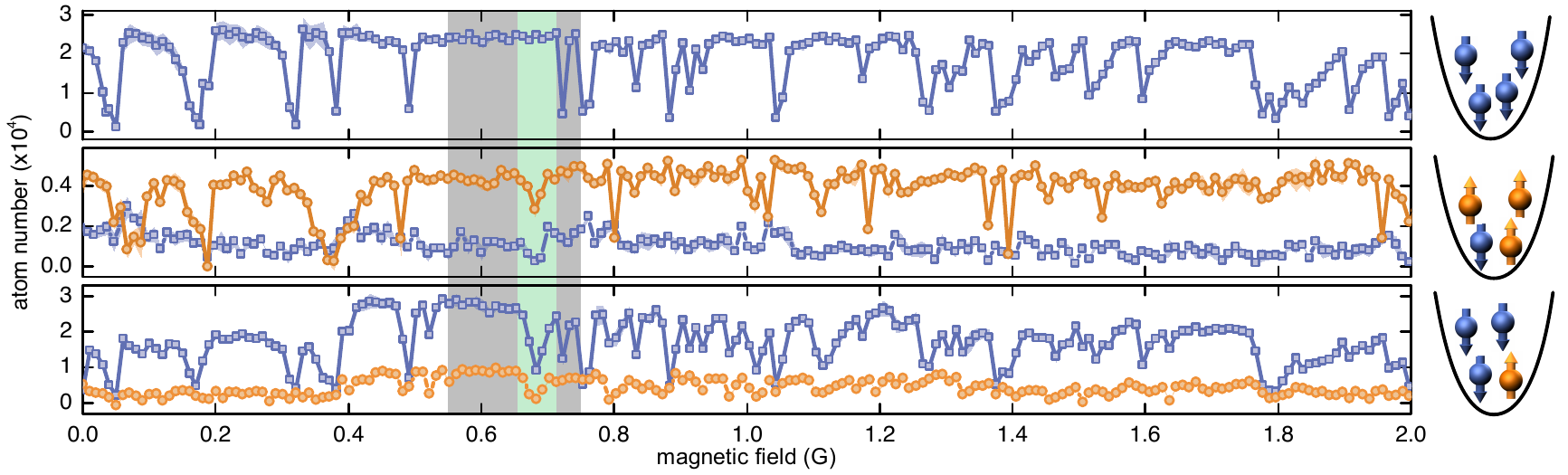}%
	\caption{\label{fig:largeFB} Feshbach spectroscopy of a two-component spin mixture in an ODT (without the lattice-protection technique) for different population imbalances $\delta$:  $\delta=1$ (upper panel), $-0.54$ (middle panel), $0.4$ (lower panel). While the measurements for the upper and lower panel are performed in $\HC$ with $\bar{\nu}=231(3)\,\rm Hz$, the data of the middle panel is measured in $\mopa$ with $\bar{\nu}=59(1)\,\rm Hz$. Due to the finite resolution of the scans of $10\,\rm mG$ it is possible that narrower FRs are not resolved. The grey shading shows the magnetic field region studied in the main manuscript with the green shading indicating the comparatively broad interspin FR. {Each data point is the mean of two repetitions.} The shading around the data points indicates statistical uncertainties, which are often smaller than the data points.}
\end{figure*}

We analyze the three sets of data to extract the spin nature of the individual FRs. For several FRs, the entrance spin channel can be easily identified. In addition, we also observe overlapping FRs. Here, an exact assignment requires a high-resolution magnetic-field scan and our lattice-protection technique; see main text. Among the forest of FRs recorded in the two-component mixture, we observe a promising interspin FR at about $\unit[700]{mG}$, which remains rather isolated from other homo-spin FRs; see green shading in Fig.\,S3.

As a second step, we focus on the magnetic-field region around $\unit[700]{mG}$ in which the promising interspin Feshbach resonance has been identified and perform high-resolution Feshbach spectroscopy, taking advantage of the lattice-preparation scheme, as described in the main text. The  lattice-protection technique is very powerful in removing technical broadening and artificial asymmetry of the loss peaks, as it clearly appears from a comparison between the atom-number traces recorded with ODT-preparation (Fig.\,S3) and lattice-preparation schemes (Fig.\,3). We perform the measurements for Fig.\,3 as follows. We prepare a spin mixture in the lattice at high $B$ as described above and sub-sequentially ramp the field to the desired value within $\unit[10]{ms}$. After letting the $B$-field stabilize for about $\unit[100]{ms}$, the dipole trap beams are ramped up within $\unit[10]{ms}$ and we unload the atoms from the lattice back into the $\mopa$ within $\unit[150]{ms}$. At this stage, the sample contains $N\approx 1.6\times10^4$ atoms at $T\approx 0.3\,T_{\rm F}$, almost independent of $\delta$, and the trap frequencies are $(\nu_{\perp},\nu_{\parallel},\nu_{z})=(111.6(2), 35(1), 169.4(6))\,\rm Hz$. We then record the spin population after a holding time of $\unit[500]{ms}$. For each magnetic field value, the measurement is repeated between two to four times and the average is reported in Fig.\,3.

 For all the above described measurements, we note that the observed atom losses can be mainly attributed to resonant three-body recombination collisions in the short-range potential. Inelastic two-body losses driven by the spin-non-conserving dipolar interactions are, in principle, also energetically allowed since $\seven$ atoms are in an excited Zeeman state \cite{Chin2010fri}. However, we do not expect this process to be enhanced at resonance.

\subsection{Modulation spectroscopy with a fermionic spin mixture in the lattice}
To measure the scattering length between two spin states of $\Er$, we rely on a method similar to the one that we have already successfully implemented with $^{168}$Er~\cite{Baier2016ebh} and $^{166}$Er~\cite{Chomaz2016qfd}. It is based on the measurement of the onsite interaction energy of two atoms in a deep optical lattice. Here, after preparing a spin mixture of $\nine$ and $\seven$ in the lattice we drive particle-hole excitations of neighboring atoms 
by a resonant modulation of the horizontal lattice depths $s_{x,y}$. Note that for neighboring identical particles no single-band excitation will be observed due to the Pauli exclusion principle and only neighboring atoms in different spin states can be excited, allowing to effectively only probe the interspin onsite interaction $U_{\down\up}$. In our experiment, we probe the doublon creation via the resulting increase in atom loss. We speculate that onsite dipolar relaxation is responsible for the observed loss~\cite{dePaz2013rdo}.  We note that a distinct and convenient method to measure double occupancies has been demonstrated using the coupling to a third spin state~\cite{Jordens2008ami}. Close to a molecular state of the original doublon components, the third spin state features a smaller interspin onsite energy and can thus be used to detect an initial double occupancy. Yet, such a method remains to be explored in our system.

In our experiment, we typically modulate the lattice depth for $\unit[1]{s}$ with a sine function with a peak-to-peak amplitude of $30\%$ and a frequency  $\nu_{\rm mod}$. Maximum loss occurs when $\nu_{\rm mod}$ reaches the resonance condition $\nu_{\rm res}=U_{\down\up}/h$ (see Fig.\,S4). 
\begin{figure}
	\includegraphics[width=1\linewidth]{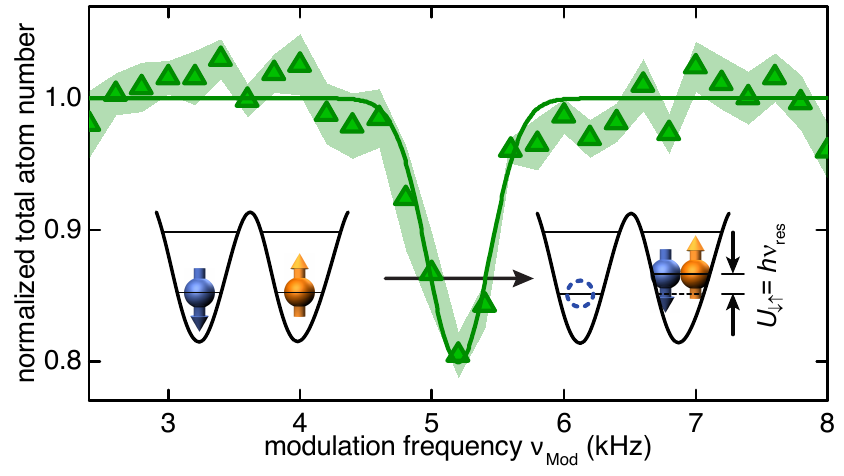}%
	\caption{\label{fig:ResMod}
    Exemplary modulation spectroscopy measurement with a spin mixture of $\nine$ and $\seven$ in the deep lattice at $B=650\,\rm mG$. The resonance condition determines $\nu_{\rm res}$, which is related to the onsite energy $U_{\down\up}$ (cartoon).
	}
\end{figure}
Following our previous work~\cite{Baier2016ebh}, the onsite energy $U_{\down\up}$ consists out of two contributions: the contact interaction $U_{\rm c}$ 
\begin{equation*}
U_{\rm c}=\frac{4\pi \hbar a_{\down\up}}{m_{\mathrm{Er}}}\int d\mathbf{r}\left\vert \phi(\mathbf{r})\right\vert ^{4},
\end{equation*}%
and the DDI $U_{\rm dd}$
\begin{equation*}
U_{\rm dd}=\frac{\mu _{0}\mu_{\down}\mu_{\up}}{4\pi }\int d\mathbf{r}\int d\mathbf{r^{\prime
}}\left\vert \phi(\mathbf{r})\right\vert ^{2}\frac{1-3\cos ^{2}\theta _{%
\mathbf{r}-\mathbf{r}^{\prime }}}{|\mathbf{r}-\mathbf{r}^{\prime }|^{3}}%
\left\vert \phi (\mathbf{r}^{\prime })\right\vert ^{2}.
\end{equation*}%
Here, $\phi(\mathbf{r})$ denotes the onsite Wannier function, $|\mathbf{r-r^\prime}|$ is the interatomic distance and $\theta _{\mathbf{r-r^\prime}}$ corresponds to the angle between the polarization axis of the two dipoles with respect to their interparticle axis. The contact part depends on the interspin scattering length $a_{\down\up}$, the reduced Plank constant $\hbar$, and the mass $m_{\mathrm{Er}}$ of a $\Er$ atom, while the DDI part is proportional to the vacuum permeability $\mu_0$ and to the magnetic moments of the two spin states $\mu_{\down}$ and $\mu_{\up}$.
The contributions of nearest-neighbor interactions are minor and therefore neglected.

Both, the strength and the sign of $U_{\rm dd}$ strongly depend on the dipole orientation and the anisotropy of the onsite Wannier function. As specified in our earlier work~\cite{Baier2016ebh}, we define the aspect ratio, $\mathrm{AR}$, associated to the Wannier function by the ratio of the onsite harmonic oscillator lengths perpendicular and in the $xy$-plane, $\mathrm{AR}=l_z/l_{x,y}$. Note that, in a deep lattice, the onsite harmonic oscillator lengths match $l_i=d_i/(\pi s_i^{1/4})$ for $i\in \{x,y,z\}$.
For our typical lattice parameters we find $\mathrm{AR}>1$ and hence $U_{\rm dd}$ can be tuned by rotating the atomic dipole. In particular, $U_{\rm dd}$ is negative (positive) for a dipole orientation out of (in) the $xy$-plane. 

{In the experiment, we use both our precise control and our exact knowledge of $U_{\rm dd}$ to determine not only the amplitude but also the sign of the scattering length $a_{\down\up}$. For a given magnetic field and a given lattice configuration, we repeat our modulation spectroscopy measurements for two different dipole orientations: (i) when oriented along $z$, we extract the total onsite energy $|U_{\down\up}^z|$ while knowing the dipolar contribution $U_{\rm dd}^z$, (ii) when oriented in the $xy$-plane, we extract $|U_{\down\up}^{xy}|$ while knowing the dipolar contribution $U_{\rm dd}^{xy}$. This yields the two indepedent and incommensurate relations: $|U_{\down\up}^z|=|U_c +U_{\rm dd}^z|$ and $|U_{\down\up}^{xy}|=|U_c +U_{\rm dd}^{xy}|$. Their combination gives access to both the magnitude and the sign of $U_c$, and thus of $a_{\down\up}$ as reported in Fig.\,4(a).}


As a final test of our method we study the dependence of the onsite energy as a function of the lattice depth $s_{z}$ (Fig.\,S5). Here, we fix the magnetic field, oriented along $z$, to $650\,\rm mG$ and vary the depth of the $z$ lattice. We repeat the modulation spectroscopy for different values of $s_z$ and extract $\nu_{\rm res}$ for each measurement. A comparison to our theoretical model with $a_{\down\up}$ being the only free parameter shows a good agreement, confirming the validity of our modulation spectroscopy technique. Here, the fit gives a value for $a_{\down\up}$ of $225(2)\,\rm a_0$ matching the value extracted from an independent analysis of the individual lattice configurations as reported in Fig.\,4(a) and giving {$a_{\down\up}=225(4)\,\rm a_0$}.
\begin{figure}
	\includegraphics[width=1\linewidth]{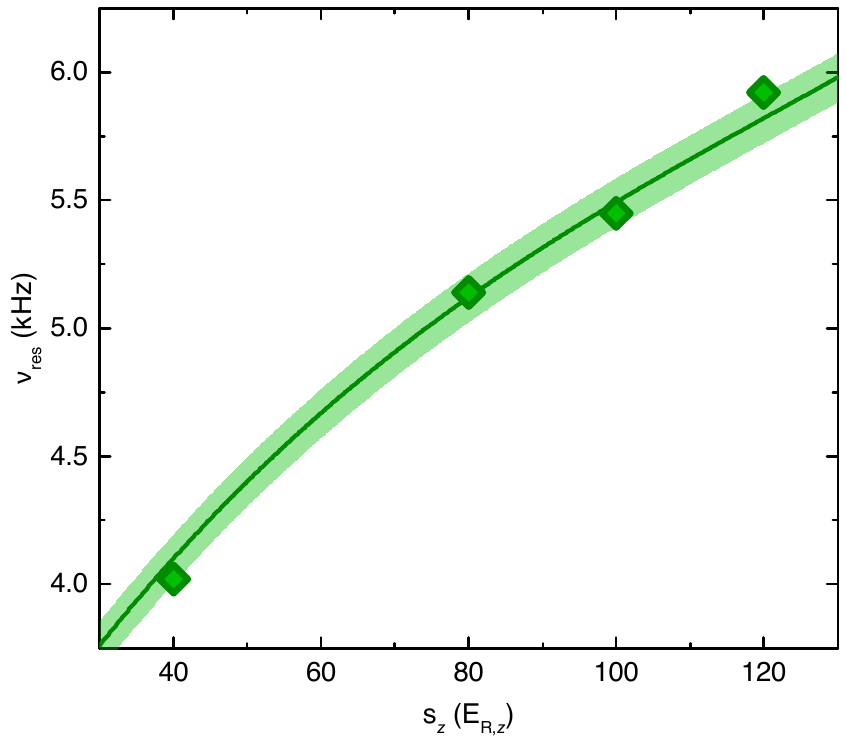}%
	\caption{\label{fig:ModSpect}
    Modulation resonance $\nu_{\rm res}$ as a function of the vertical lattice power $s_z$ for $s_{x,y}=20$ at $B=650\,\rm mG$. The solid line shows a fit with our theory to extract the scattering length $a_{\down\up}$. The shaded region accounts for the systematic uncertainty of the scattering length of $\pm 4\,\rm a_0$ at $0.65\,\rm G$, which results from our magnetic field fluctuations of $\pm 1\,\rm mG$.
	}
\end{figure}

The data presented in Fig.\,4(a) shows the mean of the different experimental datasets, taken with different lattice parameters, for a given magnetic field $B$. Table 1 summarizes all lattice parameters used in the experiment as well as the expected values of $U_{\rm c}$ for $a_{\down\up}=100\,a_0$, denoted $U_{\rm c}^{(100)}$, and of $U_{\rm dd}^z$ from our theoretical model. $U_{\rm c}$ being proportional to $a_{\down\up}$ and $U_{\rm dd}$ depending only on the lattice parameters, the interspin scattering length can be evaluated from a given measurement of $U_{\down\up}$ via $a_{\down\up}/a_0=(U_{\down\up}-U_{\rm dd}^z)\times 100/U_{\rm c}^{(100)}$.
\begin{table}
		\begin{ruledtabular}
		\begin{tabular}{l c c r}
			($s_x,s_y,s_z$)& $\rm AR$ & $U_{\rm c}^{(100)}/h\,\rm (Hz)$ & $U_{\rm dd}^z/h\,\rm (Hz)$\\	\hline
			($20,20,40$)& $1.68$ & $2029$ & $-441$\\	
			($20,20,60$)& $1.52$ & $2263$ & $-396$\\	
			($20,20,80$)& $1.41$ & $2443$ & $-350$\\		
			($20,20,100$)& $1.34$ & $2590$ & $-307$\\	
			($20,20,120$)& $1.28$ & $2717$ & $-265$\\	
			($15,15,80$)& $1.32$ & $2068$ & $-223$\\	
			($22,22,80$)& $1.45$ & $2578$ & $-399$\\	
		\end{tabular}
		\end{ruledtabular}
	\caption{Lattice parameters for the determination of $a_{\down\up}$ (Fig.\,4(a)). The lattice depths ($s_x,s_y,s_z$) define the onsite Wannier function $\rm AR$. From our theoretical model we evaluate the onsite energy contributions $U_{\rm c}^{(100)}$ and $U_{\rm dd}^z$ for an interspin scattering length of $a_{\down\up}=100\,\rm a_0$. Here, the dipoles are oriented along $z$. This values are used to extract the interspin scattering length from the measured total onsite energy $U_{\down\up}$.}
	\label{table}	
\end{table}

\section{{Scattering-length tunability and magnetic-field stability.}}

{A precise control of the magnetic-field value is crucial for tuning the interaction strength in the spin mixture. For the FR of Fig.\,4(a), a width of $\Delta=58(6)$\,mG and relative strength $a_{\rm bg}/R^*=0.1$ have been estimated. Hence, reaching $a_{\down \up}(B)=R^*$ requires to sit $\approx 6$\,mG away from the resonance pole. Based on an rf-spectroscopy calibration scheme, an accuracy of the order of 100\,$\mu$G and a stability of $\approx 1$mG on the magnetic-field value are estimated, corresponding to a relative stability of $10^{-3}$ at the resonance position $B_0=687$\,mG. 
Note that in alkali Fermi experiments, for which $B_0$ is typically 3 orders of magnitude larger, a much larger relative stability of $10^{-5}$ is usually required even if the FR is effectively broader.
}

\subsection{Loss spectroscopy in the ODT at the interspin FR}
For the measurements of the collisional properties of the fermionic spin mixture in the vicinity of the comparatively broad interspin FR (see Fig.\,4(b-d)), we apply the following experimental procedure.
We prepare a spin mixture with $\delta=0$ in the deep 3D lattice following the scheme detailed above, that is applying a RF-sweep at large $B$. After the application of the RF-sweep, we ramp the magnetic-field value to an intermediate lower value $B=\unit[3.99]{G}$ in 100\,ms. We then jump with the magnetic field from the later intermediate value to the desired final value and let it stabilize for $\unit[10]{ms}$. Finally, we ramp up the $\mopa$ beams in $\unit[10]{ms}$, melt the lattice down in $\unit[20]{ms}$. This shorter timescale for the lattice rampdown (compared, e.\,g.\,, to the Feshbach spectroscopy measurements, see above) is chosen to avoid significant losses to happen already at this stage. The final trap frequencies in the $\mopa$ are $(\nu_{\perp},\nu_{\parallel},\nu_{z})=(111.6(2), 35(1), 169.4(6))\,\rm Hz$. For this trap, the typical atom numbers recorded (see Fig.\,4(b-c)) correspond to the Fermi energy $E_{\mathrm{F}}\approx k_{\mathrm{B}}\times\unit[150]{nK}$ for each spin component, which in turn gives a Fermi wave vector $k_{\rm F}\approx\sqrt{2 m_{\mathrm{Er}} E_{\mathrm{F}}}/\hbar= 1\times 10^7\,\mathrm{m^{-1}}$. We then hold the two-component mixture in the $\mopa$ for a variable holding time, $t$, at the selected $B$-field and ultimately record the spin populations via Stern-Gerlach imaging. 

We record the atom number decay with $t$ for various magnetic fields $B$ across the FR. For each $B$ and each spin component, we extract an initial decay rate $\dot{N}/N_0$ by fitting a linear-decay function to the recorded atom number $N$, normalized to its initial value $N_0$, as a function of $t$. We fit all data for which the atom number stays above a threshold of $75\%$ of $N_0$. We checked that the extracted values of  $\dot{N}/N_0$ do not change significantly when varying this threshold between $65-85\%$. An analysis of the full data using exponential fits also yields similar decay rate values.


\begin{thebibliography}{54}%
\makeatletter
\providecommand \@ifxundefined [1]{%
 \@ifx{#1\undefined}
}%
\providecommand \@ifnum [1]{%
 \ifnum #1\expandafter \@firstoftwo
 \else \expandafter \@secondoftwo
 \fi
}%
\providecommand \@ifx [1]{%
 \ifx #1\expandafter \@firstoftwo
 \else \expandafter \@secondoftwo
 \fi
}%
\providecommand \natexlab [1]{#1}%
\providecommand \enquote  [1]{``#1''}%
\providecommand \bibnamefont  [1]{#1}%
\providecommand \bibfnamefont [1]{#1}%
\providecommand \citenamefont [1]{#1}%
\providecommand \href@noop [0]{\@secondoftwo}%
\providecommand \href [0]{\begingroup \@sanitize@url \@href}%
\providecommand \@href[1]{\@@startlink{#1}\@@href}%
\providecommand \@@href[1]{\endgroup#1\@@endlink}%
\providecommand \@sanitize@url [0]{\catcode `\\12\catcode `\$12\catcode
  `\&12\catcode `\#12\catcode `\^12\catcode `\_12\catcode `\%12\relax}%
\providecommand \@@startlink[1]{}%
\providecommand \@@endlink[0]{}%
\providecommand \url  [0]{\begingroup\@sanitize@url \@url }%
\providecommand \@url [1]{\endgroup\@href {#1}{\urlprefix }}%
\providecommand \urlprefix  [0]{URL }%
\providecommand \Eprint [0]{\href }%
\providecommand \doibase [0]{http://dx.doi.org/}%
\providecommand \selectlanguage [0]{\@gobble}%
\providecommand \bibinfo  [0]{\@secondoftwo}%
\providecommand \bibfield  [0]{\@secondoftwo}%
\providecommand \translation [1]{[#1]}%
\providecommand \BibitemOpen [0]{}%
\providecommand \bibitemStop [0]{}%
\providecommand \bibitemNoStop [0]{.\EOS\space}%
\providecommand \EOS [0]{\spacefactor3000\relax}%
\providecommand \BibitemShut  [1]{\csname bibitem#1\endcsname}%
\let\auto@bib@innerbib\@empty
\bibitem [{\citenamefont {Griesmaier}\ \emph {et~al.}(2005)\citenamefont
  {Griesmaier}, \citenamefont {Werner}, \citenamefont {Hensler}, \citenamefont
  {Stuhler},\ and\ \citenamefont {Pfau}}]{Griesmaier2005bec}%
  \BibitemOpen
  \bibfield  {author} {\bibinfo {author} {\bibfnamefont {A.}~\bibnamefont
  {Griesmaier}}, \bibinfo {author} {\bibfnamefont {J.}~\bibnamefont {Werner}},
  \bibinfo {author} {\bibfnamefont {S.}~\bibnamefont {Hensler}}, \bibinfo
  {author} {\bibfnamefont {J.}~\bibnamefont {Stuhler}}, \ and\ \bibinfo
  {author} {\bibfnamefont {T.}~\bibnamefont {Pfau}},\ }\href {\doibase
  10.1103/PhysRevLett.94.160401} {\bibfield  {journal} {\bibinfo  {journal}
  {Phys. Rev. Lett.}\ }\textbf {\bibinfo {volume} {94}},\ \bibinfo {pages}
  {160401} (\bibinfo {year} {2005})}\BibitemShut {NoStop}%
\bibitem [{\citenamefont {Lu}\ \emph {et~al.}(2011)\citenamefont {Lu},
  \citenamefont {Burdick}, \citenamefont {Youn},\ and\ \citenamefont
  {Lev}}]{Lu2011sdb}%
  \BibitemOpen
  \bibfield  {author} {\bibinfo {author} {\bibfnamefont {M.}~\bibnamefont
  {Lu}}, \bibinfo {author} {\bibfnamefont {N.~Q.}\ \bibnamefont {Burdick}},
  \bibinfo {author} {\bibfnamefont {S.~H.}\ \bibnamefont {Youn}}, \ and\
  \bibinfo {author} {\bibfnamefont {B.~L.}\ \bibnamefont {Lev}},\ }\href
  {\doibase 10.1103/PhysRevLett.107.190401} {\bibfield  {journal} {\bibinfo
  {journal} {Phys. Rev. Lett.}\ }\textbf {\bibinfo {volume} {107}},\ \bibinfo
  {pages} {190401} (\bibinfo {year} {2011})}\BibitemShut {NoStop}%
\bibitem [{\citenamefont {Lu}\ \emph {et~al.}(2012)\citenamefont {Lu},
  \citenamefont {Burdick},\ and\ \citenamefont {Lev}}]{Lu2012qdd}%
  \BibitemOpen
  \bibfield  {author} {\bibinfo {author} {\bibfnamefont {M.}~\bibnamefont
  {Lu}}, \bibinfo {author} {\bibfnamefont {N.~Q.}\ \bibnamefont {Burdick}}, \
  and\ \bibinfo {author} {\bibfnamefont {B.~L.}\ \bibnamefont {Lev}},\ }\href
  {\doibase 10.1103/PhysRevLett.108.215301} {\bibfield  {journal} {\bibinfo
  {journal} {Phys. Rev. Lett.}\ }\textbf {\bibinfo {volume} {108}},\ \bibinfo
  {pages} {215301} (\bibinfo {year} {2012})}\BibitemShut {NoStop}%
\bibitem [{\citenamefont {Aikawa}\ \emph {et~al.}(2012)\citenamefont {Aikawa},
  \citenamefont {Frisch}, \citenamefont {Mark}, \citenamefont {Baier},
  \citenamefont {Rietzler}, \citenamefont {Grimm},\ and\ \citenamefont
  {Ferlaino}}]{Aikawa2012bec}%
  \BibitemOpen
  \bibfield  {author} {\bibinfo {author} {\bibfnamefont {K.}~\bibnamefont
  {Aikawa}}, \bibinfo {author} {\bibfnamefont {A.}~\bibnamefont {Frisch}},
  \bibinfo {author} {\bibfnamefont {M.}~\bibnamefont {Mark}}, \bibinfo {author}
  {\bibfnamefont {S.}~\bibnamefont {Baier}}, \bibinfo {author} {\bibfnamefont
  {A.}~\bibnamefont {Rietzler}}, \bibinfo {author} {\bibfnamefont
  {R.}~\bibnamefont {Grimm}}, \ and\ \bibinfo {author} {\bibfnamefont
  {F.}~\bibnamefont {Ferlaino}},\ }\href {\doibase
  10.1103/PhysRevLett.108.210401} {\bibfield  {journal} {\bibinfo  {journal}
  {Phys. Rev. Lett.}\ }\textbf {\bibinfo {volume} {108}},\ \bibinfo {pages}
  {210401} (\bibinfo {year} {2012})}\BibitemShut {NoStop}%
\bibitem [{\citenamefont {Aikawa}\ \emph
  {et~al.}(2014{\natexlab{a}})\citenamefont {Aikawa}, \citenamefont {Frisch},
  \citenamefont {Mark}, \citenamefont {Baier}, \citenamefont {Grimm},\ and\
  \citenamefont {Ferlaino}}]{Aikawa2014rfd}%
  \BibitemOpen
  \bibfield  {author} {\bibinfo {author} {\bibfnamefont {K.}~\bibnamefont
  {Aikawa}}, \bibinfo {author} {\bibfnamefont {A.}~\bibnamefont {Frisch}},
  \bibinfo {author} {\bibfnamefont {M.}~\bibnamefont {Mark}}, \bibinfo {author}
  {\bibfnamefont {S.}~\bibnamefont {Baier}}, \bibinfo {author} {\bibfnamefont
  {R.}~\bibnamefont {Grimm}}, \ and\ \bibinfo {author} {\bibfnamefont
  {F.}~\bibnamefont {Ferlaino}},\ }\href {\doibase
  10.1103/PhysRevLett.112.010404} {\bibfield  {journal} {\bibinfo  {journal}
  {Phys. Rev. Lett.}\ }\textbf {\bibinfo {volume} {112}},\ \bibinfo {pages}
  {010404} (\bibinfo {year} {2014}{\natexlab{a}})}\BibitemShut {NoStop}%
\bibitem [{\citenamefont {Naylor}\ \emph {et~al.}(2015)\citenamefont {Naylor},
  \citenamefont {Reigue}, \citenamefont {Mar\'echal}, \citenamefont {Gorceix},
  \citenamefont {Laburthe-Tolra},\ and\ \citenamefont
  {Vernac}}]{Naylor2015cdf}%
  \BibitemOpen
  \bibfield  {author} {\bibinfo {author} {\bibfnamefont {B.}~\bibnamefont
  {Naylor}}, \bibinfo {author} {\bibfnamefont {A.}~\bibnamefont {Reigue}},
  \bibinfo {author} {\bibfnamefont {E.}~\bibnamefont {Mar\'echal}}, \bibinfo
  {author} {\bibfnamefont {O.}~\bibnamefont {Gorceix}}, \bibinfo {author}
  {\bibfnamefont {B.}~\bibnamefont {Laburthe-Tolra}}, \ and\ \bibinfo {author}
  {\bibfnamefont {L.}~\bibnamefont {Vernac}},\ }\href {\doibase
  10.1103/PhysRevA.91.011603} {\bibfield  {journal} {\bibinfo  {journal} {Phys.
  Rev. A}\ }\textbf {\bibinfo {volume} {91}},\ \bibinfo {pages} {011603}
  (\bibinfo {year} {2015})}\BibitemShut {NoStop}%
\bibitem [{\citenamefont {Lahaye}\ \emph {et~al.}(2008)\citenamefont {Lahaye},
  \citenamefont {Metz}, \citenamefont {Fr\"ohlich}, \citenamefont {Koch},
  \citenamefont {Meister}, \citenamefont {Griesmaier}, \citenamefont {Pfau},
  \citenamefont {Saito}, \citenamefont {Kawaguchi},\ and\ \citenamefont
  {Ueda}}]{Lahaye2008dwc}%
  \BibitemOpen
  \bibfield  {author} {\bibinfo {author} {\bibfnamefont {T.}~\bibnamefont
  {Lahaye}}, \bibinfo {author} {\bibfnamefont {J.}~\bibnamefont {Metz}},
  \bibinfo {author} {\bibfnamefont {B.}~\bibnamefont {Fr\"ohlich}}, \bibinfo
  {author} {\bibfnamefont {T.}~\bibnamefont {Koch}}, \bibinfo {author}
  {\bibfnamefont {M.}~\bibnamefont {Meister}}, \bibinfo {author} {\bibfnamefont
  {A.}~\bibnamefont {Griesmaier}}, \bibinfo {author} {\bibfnamefont
  {T.}~\bibnamefont {Pfau}}, \bibinfo {author} {\bibfnamefont {H.}~\bibnamefont
  {Saito}}, \bibinfo {author} {\bibfnamefont {Y.}~\bibnamefont {Kawaguchi}}, \
  and\ \bibinfo {author} {\bibfnamefont {M.}~\bibnamefont {Ueda}},\ }\href
  {\doibase 10.1103/PhysRevLett.101.080401} {\bibfield  {journal} {\bibinfo
  {journal} {Phys. Rev. Lett.}\ }\textbf {\bibinfo {volume} {101}},\ \bibinfo
  {pages} {080401} (\bibinfo {year} {2008})}\BibitemShut {NoStop}%
\bibitem [{\citenamefont {Kadau}\ \emph {et~al.}(2016)\citenamefont {Kadau},
  \citenamefont {Schmitt}, \citenamefont {Wenzel}, \citenamefont {Wink},
  \citenamefont {Maier}, \citenamefont {Ferrier-Barbut},\ and\ \citenamefont
  {Tilman}}]{Kadau2016otr}%
  \BibitemOpen
  \bibfield  {author} {\bibinfo {author} {\bibfnamefont {H.}~\bibnamefont
  {Kadau}}, \bibinfo {author} {\bibfnamefont {M.}~\bibnamefont {Schmitt}},
  \bibinfo {author} {\bibfnamefont {M.}~\bibnamefont {Wenzel}}, \bibinfo
  {author} {\bibfnamefont {C.}~\bibnamefont {Wink}}, \bibinfo {author}
  {\bibfnamefont {T.}~\bibnamefont {Maier}}, \bibinfo {author} {\bibfnamefont
  {I.}~\bibnamefont {Ferrier-Barbut}}, \ and\ \bibinfo {author} {\bibfnamefont
  {P.}~\bibnamefont {Tilman}},\ }\href {\doibase 10.1038/nature16485}
  {\bibfield  {journal} {\bibinfo  {journal} {Nature (London)}\ }\textbf
  {\bibinfo {volume} {530}},\ \bibinfo {pages} {194} (\bibinfo {year}
  {2016})}\BibitemShut {NoStop}%
\bibitem [{\citenamefont {Chomaz}\ \emph {et~al.}(2016)\citenamefont {Chomaz},
  \citenamefont {Baier}, \citenamefont {Petter}, \citenamefont {Mark},
  \citenamefont {W\"achtler}, \citenamefont {Santos},\ and\ \citenamefont
  {Ferlaino}}]{Chomaz2016qfd}%
  \BibitemOpen
  \bibfield  {author} {\bibinfo {author} {\bibfnamefont {L.}~\bibnamefont
  {Chomaz}}, \bibinfo {author} {\bibfnamefont {S.}~\bibnamefont {Baier}},
  \bibinfo {author} {\bibfnamefont {D.}~\bibnamefont {Petter}}, \bibinfo
  {author} {\bibfnamefont {M.~J.}\ \bibnamefont {Mark}}, \bibinfo {author}
  {\bibfnamefont {F.}~\bibnamefont {W\"achtler}}, \bibinfo {author}
  {\bibfnamefont {L.}~\bibnamefont {Santos}}, \ and\ \bibinfo {author}
  {\bibfnamefont {F.}~\bibnamefont {Ferlaino}},\ }\href {\doibase
  10.1103/PhysRevX.6.041039} {\bibfield  {journal} {\bibinfo  {journal} {Phys.
  Rev. X}\ }\textbf {\bibinfo {volume} {6}},\ \bibinfo {pages} {041039}
  (\bibinfo {year} {2016})}\BibitemShut {NoStop}%
\bibitem [{\citenamefont {Schmitt}\ \emph {et~al.}(2016)\citenamefont
  {Schmitt}, \citenamefont {Wenzel}, \citenamefont {B{\"o}ttcher},
  \citenamefont {Ferrier-Barbut},\ and\ \citenamefont {Pfau}}]{Schmitt2016sbd}%
  \BibitemOpen
  \bibfield  {author} {\bibinfo {author} {\bibfnamefont {M.}~\bibnamefont
  {Schmitt}}, \bibinfo {author} {\bibfnamefont {M.}~\bibnamefont {Wenzel}},
  \bibinfo {author} {\bibfnamefont {F.}~\bibnamefont {B{\"o}ttcher}}, \bibinfo
  {author} {\bibfnamefont {I.}~\bibnamefont {Ferrier-Barbut}}, \ and\ \bibinfo
  {author} {\bibfnamefont {T.}~\bibnamefont {Pfau}},\ }\href {\doibase
  10.1038/nature20126} {\bibfield  {journal} {\bibinfo  {journal} {Nature
  (London)}\ }\textbf {\bibinfo {volume} {539}},\ \bibinfo {pages} {259}
  (\bibinfo {year} {2016})}\BibitemShut {NoStop}%
\bibitem [{\citenamefont {Chomaz}\ \emph {et~al.}(2018)\citenamefont {Chomaz},
  \citenamefont {van Bijnen}, \citenamefont {Petter}, \citenamefont {Faraoni},
  \citenamefont {Baier}, \citenamefont {Becher}, \citenamefont {Mark},
  \citenamefont {Waechtler}, \citenamefont {Santos},\ and\ \citenamefont
  {Ferlaino}}]{chomaz2017oot}%
  \BibitemOpen
  \bibfield  {author} {\bibinfo {author} {\bibfnamefont {L.}~\bibnamefont
  {Chomaz}}, \bibinfo {author} {\bibfnamefont {R.~M.~W.}\ \bibnamefont {van
  Bijnen}}, \bibinfo {author} {\bibfnamefont {D.}~\bibnamefont {Petter}},
  \bibinfo {author} {\bibfnamefont {G.}~\bibnamefont {Faraoni}}, \bibinfo
  {author} {\bibfnamefont {S.}~\bibnamefont {Baier}}, \bibinfo {author}
  {\bibfnamefont {J.~H.}\ \bibnamefont {Becher}}, \bibinfo {author}
  {\bibfnamefont {M.~J.}\ \bibnamefont {Mark}}, \bibinfo {author}
  {\bibfnamefont {F.}~\bibnamefont {Waechtler}}, \bibinfo {author}
  {\bibfnamefont {L.}~\bibnamefont {Santos}}, \ and\ \bibinfo {author}
  {\bibfnamefont {F.}~\bibnamefont {Ferlaino}},\ }\href {\doibase
  10.1038/s41567-018-0054-7} {\bibfield  {journal} {\bibinfo  {journal} {Nature
  Physics}\ } (\bibinfo {year} {2018}),\ 10.1038/s41567-018-0054-7}\BibitemShut
  {NoStop}%
\bibitem [{\citenamefont {Aikawa}\ \emph
  {et~al.}(2014{\natexlab{b}})\citenamefont {Aikawa}, \citenamefont {Baier},
  \citenamefont {Frisch}, \citenamefont {Mark}, \citenamefont {Ravensbergen},\
  and\ \citenamefont {Ferlaino}}]{Aikawa2014oof}%
  \BibitemOpen
  \bibfield  {author} {\bibinfo {author} {\bibfnamefont {K.}~\bibnamefont
  {Aikawa}}, \bibinfo {author} {\bibfnamefont {S.}~\bibnamefont {Baier}},
  \bibinfo {author} {\bibfnamefont {A.}~\bibnamefont {Frisch}}, \bibinfo
  {author} {\bibfnamefont {M.}~\bibnamefont {Mark}}, \bibinfo {author}
  {\bibfnamefont {C.}~\bibnamefont {Ravensbergen}}, \ and\ \bibinfo {author}
  {\bibfnamefont {F.}~\bibnamefont {Ferlaino}},\ }\href {\doibase
  10.1126/science.1255259} {\bibfield  {journal} {\bibinfo  {journal}
  {Science}\ }\textbf {\bibinfo {volume} {345}},\ \bibinfo {pages} {1484}
  (\bibinfo {year} {2014}{\natexlab{b}})}\BibitemShut {NoStop}%
\bibitem [{\citenamefont {Hensler}\ \emph {et~al.}(2003)\citenamefont
  {Hensler}, \citenamefont {Werner}, \citenamefont {Griesmaier}, \citenamefont
  {Schmidt}, \citenamefont {G{\"o}rlitz}, \citenamefont {Pfau}, \citenamefont
  {Giovanazzi},\ and\ \citenamefont {Rza{\.{z}}ewski}}]{Hensler2003dri}%
  \BibitemOpen
  \bibfield  {author} {\bibinfo {author} {\bibfnamefont {S.}~\bibnamefont
  {Hensler}}, \bibinfo {author} {\bibfnamefont {J.}~\bibnamefont {Werner}},
  \bibinfo {author} {\bibfnamefont {A.}~\bibnamefont {Griesmaier}}, \bibinfo
  {author} {\bibfnamefont {P.}~\bibnamefont {Schmidt}}, \bibinfo {author}
  {\bibfnamefont {A.}~\bibnamefont {G{\"o}rlitz}}, \bibinfo {author}
  {\bibfnamefont {T.}~\bibnamefont {Pfau}}, \bibinfo {author} {\bibfnamefont
  {S.}~\bibnamefont {Giovanazzi}}, \ and\ \bibinfo {author} {\bibfnamefont
  {K.}~\bibnamefont {Rza{\.{z}}ewski}},\ }\href {\doibase
  10.1007/s00340-003-1334-0} {\bibfield  {journal} {\bibinfo  {journal}
  {Applied Physics B}\ }\textbf {\bibinfo {volume} {77}},\ \bibinfo {pages}
  {765} (\bibinfo {year} {2003})}\BibitemShut {NoStop}%
\bibitem [{\citenamefont {Pasquiou}\ \emph {et~al.}(2011)\citenamefont
  {Pasquiou}, \citenamefont {Mar\'echal}, \citenamefont {Bismut}, \citenamefont
  {Pedri}, \citenamefont {Vernac}, \citenamefont {Gorceix},\ and\ \citenamefont
  {Laburthe-Tolra}}]{Pasquiou2011sdo}%
  \BibitemOpen
  \bibfield  {author} {\bibinfo {author} {\bibfnamefont {B.}~\bibnamefont
  {Pasquiou}}, \bibinfo {author} {\bibfnamefont {E.}~\bibnamefont
  {Mar\'echal}}, \bibinfo {author} {\bibfnamefont {G.}~\bibnamefont {Bismut}},
  \bibinfo {author} {\bibfnamefont {P.}~\bibnamefont {Pedri}}, \bibinfo
  {author} {\bibfnamefont {L.}~\bibnamefont {Vernac}}, \bibinfo {author}
  {\bibfnamefont {O.}~\bibnamefont {Gorceix}}, \ and\ \bibinfo {author}
  {\bibfnamefont {B.}~\bibnamefont {Laburthe-Tolra}},\ }\href {\doibase
  10.1103/PhysRevLett.106.255303} {\bibfield  {journal} {\bibinfo  {journal}
  {Phys. Rev. Lett.}\ }\textbf {\bibinfo {volume} {106}},\ \bibinfo {pages}
  {255303} (\bibinfo {year} {2011})}\BibitemShut {NoStop}%
\bibitem [{\citenamefont {de~Paz}\ \emph
  {et~al.}(2013{\natexlab{a}})\citenamefont {de~Paz}, \citenamefont {Sharma},
  \citenamefont {Chotia}, \citenamefont {Mar\'echal}, \citenamefont {Huckans},
  \citenamefont {Pedri}, \citenamefont {Santos}, \citenamefont {Gorceix},
  \citenamefont {Vernac},\ and\ \citenamefont {Laburthe-Tolra}}]{dePaz2013nqm}%
  \BibitemOpen
  \bibfield  {author} {\bibinfo {author} {\bibfnamefont {A.}~\bibnamefont
  {de~Paz}}, \bibinfo {author} {\bibfnamefont {A.}~\bibnamefont {Sharma}},
  \bibinfo {author} {\bibfnamefont {A.}~\bibnamefont {Chotia}}, \bibinfo
  {author} {\bibfnamefont {E.}~\bibnamefont {Mar\'echal}}, \bibinfo {author}
  {\bibfnamefont {J.~H.}\ \bibnamefont {Huckans}}, \bibinfo {author}
  {\bibfnamefont {P.}~\bibnamefont {Pedri}}, \bibinfo {author} {\bibfnamefont
  {L.}~\bibnamefont {Santos}}, \bibinfo {author} {\bibfnamefont
  {O.}~\bibnamefont {Gorceix}}, \bibinfo {author} {\bibfnamefont
  {L.}~\bibnamefont {Vernac}}, \ and\ \bibinfo {author} {\bibfnamefont
  {B.}~\bibnamefont {Laburthe-Tolra}},\ }\href {\doibase
  10.1103/PhysRevLett.111.185305} {\bibfield  {journal} {\bibinfo  {journal}
  {Phys. Rev. Lett.}\ }\textbf {\bibinfo {volume} {111}},\ \bibinfo {pages}
  {185305} (\bibinfo {year} {2013}{\natexlab{a}})}\BibitemShut {NoStop}%
\bibitem [{\citenamefont {Lepoutre}\ \emph {et~al.}(2018)\citenamefont
  {Lepoutre}, \citenamefont {Schachenmayer}, \citenamefont {Gabardos},
  \citenamefont {Zhu}, \citenamefont {Naylor}, \citenamefont {Marechal},
  \citenamefont {Gorceix}, \citenamefont {Rey}, \citenamefont {Vernac},\ and\
  \citenamefont {Laburthe-Tolra}}]{Lepoutre2018eoo}%
  \BibitemOpen
  \bibfield  {author} {\bibinfo {author} {\bibfnamefont {S.}~\bibnamefont
  {Lepoutre}}, \bibinfo {author} {\bibfnamefont {J.}~\bibnamefont
  {Schachenmayer}}, \bibinfo {author} {\bibfnamefont {L.}~\bibnamefont
  {Gabardos}}, \bibinfo {author} {\bibfnamefont {B.}~\bibnamefont {Zhu}},
  \bibinfo {author} {\bibfnamefont {B.}~\bibnamefont {Naylor}}, \bibinfo
  {author} {\bibfnamefont {E.}~\bibnamefont {Marechal}}, \bibinfo {author}
  {\bibfnamefont {O.}~\bibnamefont {Gorceix}}, \bibinfo {author} {\bibfnamefont
  {A.~M.}\ \bibnamefont {Rey}}, \bibinfo {author} {\bibfnamefont
  {L.}~\bibnamefont {Vernac}}, \ and\ \bibinfo {author} {\bibfnamefont
  {B.}~\bibnamefont {Laburthe-Tolra}},\ }\href
  {https://arxiv.org/abs/1803.02628} {\bibfield  {journal} {\bibinfo  {journal}
  {ArXiv e-prints}\ } (\bibinfo {year} {2018})},\ \Eprint
  {http://arxiv.org/abs/1803.02628} {arXiv:1803.02628 [cond-mat.quant-gas]}
  \BibitemShut {NoStop}%
\bibitem [{\citenamefont {Burdick}\ \emph {et~al.}(2015)\citenamefont
  {Burdick}, \citenamefont {Baumann}, \citenamefont {Tang}, \citenamefont
  {Lu},\ and\ \citenamefont {Lev}}]{Burdick2015fso}%
  \BibitemOpen
  \bibfield  {author} {\bibinfo {author} {\bibfnamefont {N.~Q.}\ \bibnamefont
  {Burdick}}, \bibinfo {author} {\bibfnamefont {K.}~\bibnamefont {Baumann}},
  \bibinfo {author} {\bibfnamefont {Y.}~\bibnamefont {Tang}}, \bibinfo {author}
  {\bibfnamefont {M.}~\bibnamefont {Lu}}, \ and\ \bibinfo {author}
  {\bibfnamefont {B.~L.}\ \bibnamefont {Lev}},\ }\href {\doibase
  10.1103/PhysRevLett.114.023201} {\bibfield  {journal} {\bibinfo  {journal}
  {Phys. Rev. Lett.}\ }\textbf {\bibinfo {volume} {114}},\ \bibinfo {pages}
  {023201} (\bibinfo {year} {2015})}\BibitemShut {NoStop}%
\bibitem [{\citenamefont {Burdick}\ \emph {et~al.}(2016)\citenamefont
  {Burdick}, \citenamefont {Tang},\ and\ \citenamefont {Lev}}]{Burdick2016lls}%
  \BibitemOpen
  \bibfield  {author} {\bibinfo {author} {\bibfnamefont {N.~Q.}\ \bibnamefont
  {Burdick}}, \bibinfo {author} {\bibfnamefont {Y.}~\bibnamefont {Tang}}, \
  and\ \bibinfo {author} {\bibfnamefont {B.~L.}\ \bibnamefont {Lev}},\ }\href
  {\doibase 10.1103/PhysRevX.6.031022} {\bibfield  {journal} {\bibinfo
  {journal} {Phys. Rev. X}\ }\textbf {\bibinfo {volume} {6}},\ \bibinfo {pages}
  {031022} (\bibinfo {year} {2016})}\BibitemShut {NoStop}%
\bibitem [{\citenamefont {Baranov}\ \emph {et~al.}(2012)\citenamefont
  {Baranov}, \citenamefont {Dalmonte}, \citenamefont {Pupillo},\ and\
  \citenamefont {Zoller}}]{Baranov2012cmt}%
  \BibitemOpen
  \bibfield  {author} {\bibinfo {author} {\bibfnamefont {M.~A.}\ \bibnamefont
  {Baranov}}, \bibinfo {author} {\bibfnamefont {M.}~\bibnamefont {Dalmonte}},
  \bibinfo {author} {\bibfnamefont {G.}~\bibnamefont {Pupillo}}, \ and\
  \bibinfo {author} {\bibfnamefont {P.}~\bibnamefont {Zoller}},\ }\href
  {\doibase 10.1021/cr2003568} {\bibfield  {journal} {\bibinfo  {journal}
  {Chem. Rev.}\ }\textbf {\bibinfo {volume} {112}},\ \bibinfo {pages} {5012}
  (\bibinfo {year} {2012})}\BibitemShut {NoStop}%
\bibitem [{\citenamefont {Bloch}\ \emph {et~al.}(2008)\citenamefont {Bloch},
  \citenamefont {Dalibard},\ and\ \citenamefont {Zwerger}}]{Bloch2008mbp}%
  \BibitemOpen
  \bibfield  {author} {\bibinfo {author} {\bibfnamefont {I.}~\bibnamefont
  {Bloch}}, \bibinfo {author} {\bibfnamefont {J.}~\bibnamefont {Dalibard}}, \
  and\ \bibinfo {author} {\bibfnamefont {W.}~\bibnamefont {Zwerger}},\ }\href
  {\doibase 10.1103/RevModPhys.80.885} {\bibfield  {journal} {\bibinfo
  {journal} {Rev. Mod. Phys.}\ }\textbf {\bibinfo {volume} {80}},\ \bibinfo
  {pages} {885} (\bibinfo {year} {2008})}\BibitemShut {NoStop}%
\bibitem [{\citenamefont {Stamper-Kurn}\ and\ \citenamefont
  {Ueda}(2013)}]{StamperKurn2013sbg}%
  \BibitemOpen
  \bibfield  {author} {\bibinfo {author} {\bibfnamefont {D.~M.}\ \bibnamefont
  {Stamper-Kurn}}\ and\ \bibinfo {author} {\bibfnamefont {M.}~\bibnamefont
  {Ueda}},\ }\href {\doibase 10.1103/RevModPhys.85.1191} {\bibfield  {journal}
  {\bibinfo  {journal} {Rev. Mod. Phys.}\ }\textbf {\bibinfo {volume} {85}},\
  \bibinfo {pages} {1191} (\bibinfo {year} {2013})}\BibitemShut {NoStop}%
\bibitem [{\citenamefont {Dutta}\ \emph {et~al.}(2015)\citenamefont {Dutta},
  \citenamefont {Gajda}, \citenamefont {Hauke}, \citenamefont {Lewenstein},
  \citenamefont {L\"uhmann}, \citenamefont {Malomed}, \citenamefont
  {Sowi\'{n}ski},\ and\ \citenamefont {Zakrzewski}}]{Dutta2015}%
  \BibitemOpen
  \bibfield  {author} {\bibinfo {author} {\bibfnamefont {O.}~\bibnamefont
  {Dutta}}, \bibinfo {author} {\bibfnamefont {M.}~\bibnamefont {Gajda}},
  \bibinfo {author} {\bibfnamefont {P.}~\bibnamefont {Hauke}}, \bibinfo
  {author} {\bibfnamefont {M.}~\bibnamefont {Lewenstein}}, \bibinfo {author}
  {\bibfnamefont {D.-S.}\ \bibnamefont {L\"uhmann}}, \bibinfo {author}
  {\bibfnamefont {B.~A.}\ \bibnamefont {Malomed}}, \bibinfo {author}
  {\bibfnamefont {T.}~\bibnamefont {Sowi\'{n}ski}}, \ and\ \bibinfo {author}
  {\bibfnamefont {J.}~\bibnamefont {Zakrzewski}},\ }\href
  {http://stacks.iop.org/0034-4885/78/i=6/a=066001} {\bibfield  {journal}
  {\bibinfo  {journal} {Reports on Progress in Physics}\ }\textbf {\bibinfo
  {volume} {78}},\ \bibinfo {pages} {066001} (\bibinfo {year}
  {2015})}\BibitemShut {NoStop}%
\bibitem [{\citenamefont {Frisch}\ \emph {et~al.}(2014)\citenamefont {Frisch},
  \citenamefont {Mark}, \citenamefont {Aikawa}, \citenamefont {Ferlaino},
  \citenamefont {Bohn}, \citenamefont {Makrides}, \citenamefont {Petrov},\ and\
  \citenamefont {Kotochigova}}]{Frisch2014qci}%
  \BibitemOpen
  \bibfield  {author} {\bibinfo {author} {\bibfnamefont {A.}~\bibnamefont
  {Frisch}}, \bibinfo {author} {\bibfnamefont {M.}~\bibnamefont {Mark}},
  \bibinfo {author} {\bibfnamefont {K.}~\bibnamefont {Aikawa}}, \bibinfo
  {author} {\bibfnamefont {F.}~\bibnamefont {Ferlaino}}, \bibinfo {author}
  {\bibfnamefont {J.}~\bibnamefont {Bohn}}, \bibinfo {author} {\bibfnamefont
  {C.}~\bibnamefont {Makrides}}, \bibinfo {author} {\bibfnamefont
  {A.}~\bibnamefont {Petrov}}, \ and\ \bibinfo {author} {\bibfnamefont
  {S.}~\bibnamefont {Kotochigova}},\ }\href {\doibase 10.1038/nature13137}
  {\bibfield  {journal} {\bibinfo  {journal} {Nature (London)}\ }\textbf
  {\bibinfo {volume} {507}},\ \bibinfo {pages} {475} (\bibinfo {year}
  {2014})}\BibitemShut {NoStop}%
\bibitem [{\citenamefont {Baumann}\ \emph {et~al.}(2014)\citenamefont
  {Baumann}, \citenamefont {Burdick}, \citenamefont {Lu},\ and\ \citenamefont
  {Lev}}]{Baumann2014ool}%
  \BibitemOpen
  \bibfield  {author} {\bibinfo {author} {\bibfnamefont {K.}~\bibnamefont
  {Baumann}}, \bibinfo {author} {\bibfnamefont {N.~Q.}\ \bibnamefont
  {Burdick}}, \bibinfo {author} {\bibfnamefont {M.}~\bibnamefont {Lu}}, \ and\
  \bibinfo {author} {\bibfnamefont {B.~L.}\ \bibnamefont {Lev}},\ }\href
  {\doibase 10.1103/PhysRevA.89.020701} {\bibfield  {journal} {\bibinfo
  {journal} {Phys. Rev. A}\ }\textbf {\bibinfo {volume} {89}},\ \bibinfo
  {pages} {020701} (\bibinfo {year} {2014})}\BibitemShut {NoStop}%
\bibitem [{\citenamefont {Maier}\ \emph {et~al.}(2015)\citenamefont {Maier},
  \citenamefont {Kadau}, \citenamefont {Schmitt}, \citenamefont {Wenzel},
  \citenamefont {Ferrier-Barbut}, \citenamefont {Pfau}, \citenamefont {Frisch},
  \citenamefont {Baier}, \citenamefont {Aikawa}, \citenamefont {Chomaz},
  \citenamefont {Mark}, \citenamefont {Ferlaino}, \citenamefont {Makrides},
  \citenamefont {Tiesinga}, \citenamefont {Petrov},\ and\ \citenamefont
  {Kotochigova}}]{Maier2015eoc}%
  \BibitemOpen
  \bibfield  {author} {\bibinfo {author} {\bibfnamefont {T.}~\bibnamefont
  {Maier}}, \bibinfo {author} {\bibfnamefont {H.}~\bibnamefont {Kadau}},
  \bibinfo {author} {\bibfnamefont {M.}~\bibnamefont {Schmitt}}, \bibinfo
  {author} {\bibfnamefont {M.}~\bibnamefont {Wenzel}}, \bibinfo {author}
  {\bibfnamefont {I.}~\bibnamefont {Ferrier-Barbut}}, \bibinfo {author}
  {\bibfnamefont {T.}~\bibnamefont {Pfau}}, \bibinfo {author} {\bibfnamefont
  {A.}~\bibnamefont {Frisch}}, \bibinfo {author} {\bibfnamefont
  {S.}~\bibnamefont {Baier}}, \bibinfo {author} {\bibfnamefont
  {K.}~\bibnamefont {Aikawa}}, \bibinfo {author} {\bibfnamefont
  {L.}~\bibnamefont {Chomaz}}, \bibinfo {author} {\bibfnamefont {M.~J.}\
  \bibnamefont {Mark}}, \bibinfo {author} {\bibfnamefont {F.}~\bibnamefont
  {Ferlaino}}, \bibinfo {author} {\bibfnamefont {C.}~\bibnamefont {Makrides}},
  \bibinfo {author} {\bibfnamefont {E.}~\bibnamefont {Tiesinga}}, \bibinfo
  {author} {\bibfnamefont {A.}~\bibnamefont {Petrov}}, \ and\ \bibinfo {author}
  {\bibfnamefont {S.}~\bibnamefont {Kotochigova}},\ }\href {\doibase
  10.1103/PhysRevX.5.041029} {\bibfield  {journal} {\bibinfo  {journal} {Phys.
  Rev. X}\ }\textbf {\bibinfo {volume} {5}},\ \bibinfo {pages} {041029}
  (\bibinfo {year} {2015})}\BibitemShut {NoStop}%
\bibitem [{\citenamefont {Schmaljohann}\ \emph {et~al.}(2004)\citenamefont
  {Schmaljohann}, \citenamefont {Erhard}, \citenamefont {Kronj\"ager},
  \citenamefont {Kottke}, \citenamefont {van Staa}, \citenamefont
  {Cacciapuoti}, \citenamefont {Arlt}, \citenamefont {Bongs},\ and\
  \citenamefont {Sengstock}}]{Schmaljohann2004dof}%
  \BibitemOpen
  \bibfield  {author} {\bibinfo {author} {\bibfnamefont {H.}~\bibnamefont
  {Schmaljohann}}, \bibinfo {author} {\bibfnamefont {M.}~\bibnamefont
  {Erhard}}, \bibinfo {author} {\bibfnamefont {J.}~\bibnamefont {Kronj\"ager}},
  \bibinfo {author} {\bibfnamefont {M.}~\bibnamefont {Kottke}}, \bibinfo
  {author} {\bibfnamefont {S.}~\bibnamefont {van Staa}}, \bibinfo {author}
  {\bibfnamefont {L.}~\bibnamefont {Cacciapuoti}}, \bibinfo {author}
  {\bibfnamefont {J.~J.}\ \bibnamefont {Arlt}}, \bibinfo {author}
  {\bibfnamefont {K.}~\bibnamefont {Bongs}}, \ and\ \bibinfo {author}
  {\bibfnamefont {K.}~\bibnamefont {Sengstock}},\ }\href {\doibase
  10.1103/PhysRevLett.92.040402} {\bibfield  {journal} {\bibinfo  {journal}
  {Phys. Rev. Lett.}\ }\textbf {\bibinfo {volume} {92}},\ \bibinfo {pages}
  {040402} (\bibinfo {year} {2004})}\BibitemShut {NoStop}%
\bibitem [{\citenamefont {Winkler}\ \emph {et~al.}(2007)\citenamefont
  {Winkler}, \citenamefont {Lang}, \citenamefont {Thalhammer}, \citenamefont
  {Straten}, \citenamefont {Grimm},\ and\ \citenamefont
  {Denschlag}}]{Winkler2007cot}%
  \BibitemOpen
  \bibfield  {author} {\bibinfo {author} {\bibfnamefont {K.}~\bibnamefont
  {Winkler}}, \bibinfo {author} {\bibfnamefont {F.}~\bibnamefont {Lang}},
  \bibinfo {author} {\bibfnamefont {G.}~\bibnamefont {Thalhammer}}, \bibinfo
  {author} {\bibfnamefont {P.~v.~d.}\ \bibnamefont {Straten}}, \bibinfo
  {author} {\bibfnamefont {R.}~\bibnamefont {Grimm}}, \ and\ \bibinfo {author}
  {\bibfnamefont {J.~H.}\ \bibnamefont {Denschlag}},\ }\href {\doibase
  10.1103/PhysRevLett.98.043201} {\bibfield  {journal} {\bibinfo  {journal}
  {Phys. Rev. Lett.}\ }\textbf {\bibinfo {volume} {98}},\ \bibinfo {pages}
  {043201} (\bibinfo {year} {2007})}\BibitemShut {NoStop}%
\bibitem [{\citenamefont {Chotia}\ \emph {et~al.}(2012)\citenamefont {Chotia},
  \citenamefont {Neyenhuis}, \citenamefont {Moses}, \citenamefont {Yan},
  \citenamefont {Covey}, \citenamefont {Foss-Feig}, \citenamefont {Rey},
  \citenamefont {Jin},\ and\ \citenamefont {Ye}}]{chotia2012lld}%
  \BibitemOpen
  \bibfield  {author} {\bibinfo {author} {\bibfnamefont {A.}~\bibnamefont
  {Chotia}}, \bibinfo {author} {\bibfnamefont {B.}~\bibnamefont {Neyenhuis}},
  \bibinfo {author} {\bibfnamefont {S.~A.}\ \bibnamefont {Moses}}, \bibinfo
  {author} {\bibfnamefont {B.}~\bibnamefont {Yan}}, \bibinfo {author}
  {\bibfnamefont {J.~P.}\ \bibnamefont {Covey}}, \bibinfo {author}
  {\bibfnamefont {M.}~\bibnamefont {Foss-Feig}}, \bibinfo {author}
  {\bibfnamefont {A.~M.}\ \bibnamefont {Rey}}, \bibinfo {author} {\bibfnamefont
  {D.~S.}\ \bibnamefont {Jin}}, \ and\ \bibinfo {author} {\bibfnamefont
  {J.}~\bibnamefont {Ye}},\ }\href {\doibase 10.1103/PhysRevLett.108.080405}
  {\bibfield  {journal} {\bibinfo  {journal} {Phys. Rev. Lett.}\ }\textbf
  {\bibinfo {volume} {108}},\ \bibinfo {pages} {080405} (\bibinfo {year}
  {2012})}\BibitemShut {NoStop}%
\bibitem [{sup()}]{suppmat}%
  \BibitemOpen
  \href@noop {} {}\bibinfo {note} {See Supplemental Material at [URL], which
  includes Refs.\,[48--54], for details on the experimental preparation and
  measurement schemes.}\BibitemShut {Stop}%
\bibitem [{\citenamefont {Baier}\ \emph {et~al.}(2016)\citenamefont {Baier},
  \citenamefont {Mark}, \citenamefont {Petter}, \citenamefont {Aikawa},
  \citenamefont {Chomaz}, \citenamefont {Cai}, \citenamefont {Baranov},
  \citenamefont {Zoller},\ and\ \citenamefont {Ferlaino}}]{Baier2016ebh}%
  \BibitemOpen
  \bibfield  {author} {\bibinfo {author} {\bibfnamefont {S.}~\bibnamefont
  {Baier}}, \bibinfo {author} {\bibfnamefont {M.~J.}\ \bibnamefont {Mark}},
  \bibinfo {author} {\bibfnamefont {D.}~\bibnamefont {Petter}}, \bibinfo
  {author} {\bibfnamefont {K.}~\bibnamefont {Aikawa}}, \bibinfo {author}
  {\bibfnamefont {L.}~\bibnamefont {Chomaz}}, \bibinfo {author} {\bibfnamefont
  {Z.}~\bibnamefont {Cai}}, \bibinfo {author} {\bibfnamefont {M.}~\bibnamefont
  {Baranov}}, \bibinfo {author} {\bibfnamefont {P.}~\bibnamefont {Zoller}}, \
  and\ \bibinfo {author} {\bibfnamefont {F.}~\bibnamefont {Ferlaino}},\ }\href
  {\doibase 10.1126/science.aac9812} {\bibfield  {journal} {\bibinfo  {journal}
  {Science}\ }\textbf {\bibinfo {volume} {352}},\ \bibinfo {pages} {201}
  (\bibinfo {year} {2016})}\BibitemShut {NoStop}%
\bibitem [{\citenamefont {Kastberg}\ \emph {et~al.}(1995)\citenamefont
  {Kastberg}, \citenamefont {Phillips}, \citenamefont {Rolston}, \citenamefont
  {Spreeuw},\ and\ \citenamefont {Jessen}}]{Kastberg1995aco}%
  \BibitemOpen
  \bibfield  {author} {\bibinfo {author} {\bibfnamefont {A.}~\bibnamefont
  {Kastberg}}, \bibinfo {author} {\bibfnamefont {W.~D.}\ \bibnamefont
  {Phillips}}, \bibinfo {author} {\bibfnamefont {S.~L.}\ \bibnamefont
  {Rolston}}, \bibinfo {author} {\bibfnamefont {R.~J.~C.}\ \bibnamefont
  {Spreeuw}}, \ and\ \bibinfo {author} {\bibfnamefont {P.~S.}\ \bibnamefont
  {Jessen}},\ }\href {\doibase 10.1103/PhysRevLett.74.1542} {\bibfield
  {journal} {\bibinfo  {journal} {Phys. Rev. Lett.}\ }\textbf {\bibinfo
  {volume} {74}},\ \bibinfo {pages} {1542} (\bibinfo {year}
  {1995})}\BibitemShut {NoStop}%
\bibitem [{\citenamefont {Will}(2011)}]{Will2011thesis}%
  \BibitemOpen
  \bibfield  {author} {\bibinfo {author} {\bibfnamefont {S.}~\bibnamefont
  {Will}},\ }\href {http://d-nb.info/1018411739} {\bibfield  {journal}
  {\bibinfo  {journal} {PhD Thesis, Johannes Gutenberg-Universit{\"a}t, Mainz}\
  } (\bibinfo {year} {2011})}\BibitemShut {NoStop}%
\bibitem [{\citenamefont {de~Paz}\ \emph
  {et~al.}(2013{\natexlab{b}})\citenamefont {de~Paz}, \citenamefont {Chotia},
  \citenamefont {Mar\'echal}, \citenamefont {Pedri}, \citenamefont {Vernac},
  \citenamefont {Gorceix},\ and\ \citenamefont
  {Laburthe-Tolra}}]{dePaz2013rdo}%
  \BibitemOpen
  \bibfield  {author} {\bibinfo {author} {\bibfnamefont {A.}~\bibnamefont
  {de~Paz}}, \bibinfo {author} {\bibfnamefont {A.}~\bibnamefont {Chotia}},
  \bibinfo {author} {\bibfnamefont {E.}~\bibnamefont {Mar\'echal}}, \bibinfo
  {author} {\bibfnamefont {P.}~\bibnamefont {Pedri}}, \bibinfo {author}
  {\bibfnamefont {L.}~\bibnamefont {Vernac}}, \bibinfo {author} {\bibfnamefont
  {O.}~\bibnamefont {Gorceix}}, \ and\ \bibinfo {author} {\bibfnamefont
  {B.}~\bibnamefont {Laburthe-Tolra}},\ }\href {\doibase
  10.1103/PhysRevA.87.051609} {\bibfield  {journal} {\bibinfo  {journal} {Phys.
  Rev. A}\ }\textbf {\bibinfo {volume} {87}},\ \bibinfo {pages} {051609}
  (\bibinfo {year} {2013}{\natexlab{b}})}\BibitemShut {NoStop}%
\bibitem [{\citenamefont {Pasquiou}\ \emph {et~al.}(2010)\citenamefont
  {Pasquiou}, \citenamefont {Bismut}, \citenamefont {Beaufils}, \citenamefont
  {Crubellier}, \citenamefont {Mar\'echal}, \citenamefont {Pedri},
  \citenamefont {Vernac}, \citenamefont {Gorceix},\ and\ \citenamefont
  {Laburthe-Tolra}}]{Pasquiou2010dir}%
  \BibitemOpen
  \bibfield  {author} {\bibinfo {author} {\bibfnamefont {B.}~\bibnamefont
  {Pasquiou}}, \bibinfo {author} {\bibfnamefont {G.}~\bibnamefont {Bismut}},
  \bibinfo {author} {\bibfnamefont {Q.}~\bibnamefont {Beaufils}}, \bibinfo
  {author} {\bibfnamefont {A.}~\bibnamefont {Crubellier}}, \bibinfo {author}
  {\bibfnamefont {E.}~\bibnamefont {Mar\'echal}}, \bibinfo {author}
  {\bibfnamefont {P.}~\bibnamefont {Pedri}}, \bibinfo {author} {\bibfnamefont
  {L.}~\bibnamefont {Vernac}}, \bibinfo {author} {\bibfnamefont
  {O.}~\bibnamefont {Gorceix}}, \ and\ \bibinfo {author} {\bibfnamefont
  {B.}~\bibnamefont {Laburthe-Tolra}},\ }\href {\doibase
  10.1103/PhysRevA.81.042716} {\bibfield  {journal} {\bibinfo  {journal} {Phys.
  Rev. A}\ }\textbf {\bibinfo {volume} {81}},\ \bibinfo {pages} {042716}
  (\bibinfo {year} {2010})}\BibitemShut {NoStop}%
\bibitem [{\citenamefont {Frisch}\ \emph {et~al.}(2015)\citenamefont {Frisch},
  \citenamefont {Mark}, \citenamefont {Aikawa}, \citenamefont {Baier},
  \citenamefont {Grimm}, \citenamefont {Petrov}, \citenamefont {Kotochigova},
  \citenamefont {Qu\'em\'ener}, \citenamefont {Lepers}, \citenamefont
  {Dulieu},\ and\ \citenamefont {Ferlaino}}]{Frisch2015upm}%
  \BibitemOpen
  \bibfield  {author} {\bibinfo {author} {\bibfnamefont {A.}~\bibnamefont
  {Frisch}}, \bibinfo {author} {\bibfnamefont {M.}~\bibnamefont {Mark}},
  \bibinfo {author} {\bibfnamefont {K.}~\bibnamefont {Aikawa}}, \bibinfo
  {author} {\bibfnamefont {S.}~\bibnamefont {Baier}}, \bibinfo {author}
  {\bibfnamefont {R.}~\bibnamefont {Grimm}}, \bibinfo {author} {\bibfnamefont
  {A.}~\bibnamefont {Petrov}}, \bibinfo {author} {\bibfnamefont
  {S.}~\bibnamefont {Kotochigova}}, \bibinfo {author} {\bibfnamefont
  {G.}~\bibnamefont {Qu\'em\'ener}}, \bibinfo {author} {\bibfnamefont
  {M.}~\bibnamefont {Lepers}}, \bibinfo {author} {\bibfnamefont
  {O.}~\bibnamefont {Dulieu}}, \ and\ \bibinfo {author} {\bibfnamefont
  {F.}~\bibnamefont {Ferlaino}},\ }\href {\doibase
  10.1103/PhysRevLett.115.203201} {\bibfield  {journal} {\bibinfo  {journal}
  {Phys. Rev. Lett.}\ }\textbf {\bibinfo {volume} {115}},\ \bibinfo {pages}
  {203201} (\bibinfo {year} {2015})}\BibitemShut {NoStop}%
\bibitem [{\citenamefont {Chin}\ \emph {et~al.}(2010)\citenamefont {Chin},
  \citenamefont {Grimm}, \citenamefont {Julienne},\ and\ \citenamefont
  {Tiesinga}}]{Chin2010fri}%
  \BibitemOpen
  \bibfield  {author} {\bibinfo {author} {\bibfnamefont {C.}~\bibnamefont
  {Chin}}, \bibinfo {author} {\bibfnamefont {R.}~\bibnamefont {Grimm}},
  \bibinfo {author} {\bibfnamefont {P.}~\bibnamefont {Julienne}}, \ and\
  \bibinfo {author} {\bibfnamefont {E.}~\bibnamefont {Tiesinga}},\ }\href
  {\doibase 10.1103/RevModPhys.82.1225} {\bibfield  {journal} {\bibinfo
  {journal} {Rev. Mod. Phys.}\ }\textbf {\bibinfo {volume} {82}},\ \bibinfo
  {pages} {1225} (\bibinfo {year} {2010})}\BibitemShut {NoStop}%
\bibitem [{\citenamefont {Gurarie}\ and\ \citenamefont
  {Radzihovsky}(2007)}]{Gurarie2007rpf}%
  \BibitemOpen
  \bibfield  {author} {\bibinfo {author} {\bibfnamefont {V.}~\bibnamefont
  {Gurarie}}\ and\ \bibinfo {author} {\bibfnamefont {L.}~\bibnamefont
  {Radzihovsky}},\ }\href {\doibase https://doi.org/10.1016/j.aop.2006.10.009}
  {\bibfield  {journal} {\bibinfo  {journal} {Annals of Physics}\ }\textbf
  {\bibinfo {volume} {322}},\ \bibinfo {pages} {2 } (\bibinfo {year} {2007})},\
  \bibinfo {note} {january Special Issue 2007}\BibitemShut {NoStop}%
\bibitem [{\citenamefont {Ho}\ \emph {et~al.}(2012)\citenamefont {Ho},
  \citenamefont {Cui},\ and\ \citenamefont {Li}}]{Ho2012art}%
  \BibitemOpen
  \bibfield  {author} {\bibinfo {author} {\bibfnamefont {T.-L.}\ \bibnamefont
  {Ho}}, \bibinfo {author} {\bibfnamefont {X.}~\bibnamefont {Cui}}, \ and\
  \bibinfo {author} {\bibfnamefont {W.}~\bibnamefont {Li}},\ }\href {\doibase
  10.1103/PhysRevLett.108.250401} {\bibfield  {journal} {\bibinfo  {journal}
  {Phys. Rev. Lett.}\ }\textbf {\bibinfo {volume} {108}},\ \bibinfo {pages}
  {250401} (\bibinfo {year} {2012})}\BibitemShut {NoStop}%
\bibitem [{\citenamefont {Hazlett}\ \emph {et~al.}(2012)\citenamefont
  {Hazlett}, \citenamefont {Zhang}, \citenamefont {Stites},\ and\ \citenamefont
  {O'Hara}}]{Hazlett2012roa}%
  \BibitemOpen
  \bibfield  {author} {\bibinfo {author} {\bibfnamefont {E.~L.}\ \bibnamefont
  {Hazlett}}, \bibinfo {author} {\bibfnamefont {Y.}~\bibnamefont {Zhang}},
  \bibinfo {author} {\bibfnamefont {R.~W.}\ \bibnamefont {Stites}}, \ and\
  \bibinfo {author} {\bibfnamefont {K.~M.}\ \bibnamefont {O'Hara}},\ }\href
  {\doibase 10.1103/PhysRevLett.108.045304} {\bibfield  {journal} {\bibinfo
  {journal} {Phys. Rev. Lett.}\ }\textbf {\bibinfo {volume} {108}},\ \bibinfo
  {pages} {045304} (\bibinfo {year} {2012})}\BibitemShut {NoStop}%
\bibitem [{\citenamefont {Inguscio}\ \emph {et~al.}(2007)\citenamefont
  {Inguscio}, \citenamefont {Ketterle},\ and\ \citenamefont
  {Salomon}}]{Varenna2007ucf}%
  \BibitemOpen
  \bibinfo {editor} {\bibfnamefont {M.}~\bibnamefont {Inguscio}}, \bibinfo
  {editor} {\bibfnamefont {W.}~\bibnamefont {Ketterle}}, \ and\ \bibinfo
  {editor} {\bibfnamefont {C.}~\bibnamefont {Salomon}},\ eds.,\ \href@noop {}
  {\emph {\bibinfo {title} {Proceedings of the International School of Physics
  "Enrico Fermi", Course CLXIV}}}\ (\bibinfo  {publisher} {IOS Press},\
  \bibinfo {address} {Amsterdam},\ \bibinfo {year} {2007})\BibitemShut
  {NoStop}%
\bibitem [{\citenamefont {Petrov}(2003)}]{Petrov2003tbp}%
  \BibitemOpen
  \bibfield  {author} {\bibinfo {author} {\bibfnamefont {D.~S.}\ \bibnamefont
  {Petrov}},\ }\href {\doibase 10.1103/PhysRevA.67.010703} {\bibfield
  {journal} {\bibinfo  {journal} {Phys. Rev. A}\ }\textbf {\bibinfo {volume}
  {67}},\ \bibinfo {pages} {010703} (\bibinfo {year} {2003})}\BibitemShut
  {NoStop}%
\bibitem [{\citenamefont {Dieckmann}\ \emph {et~al.}(2002)\citenamefont
  {Dieckmann}, \citenamefont {Stan}, \citenamefont {Gupta}, \citenamefont
  {Hadzibabic}, \citenamefont {Schunck},\ and\ \citenamefont
  {Ketterle}}]{Dieckmann2002doa}%
  \BibitemOpen
  \bibfield  {author} {\bibinfo {author} {\bibfnamefont {K.}~\bibnamefont
  {Dieckmann}}, \bibinfo {author} {\bibfnamefont {C.~A.}\ \bibnamefont {Stan}},
  \bibinfo {author} {\bibfnamefont {S.}~\bibnamefont {Gupta}}, \bibinfo
  {author} {\bibfnamefont {Z.}~\bibnamefont {Hadzibabic}}, \bibinfo {author}
  {\bibfnamefont {C.~H.}\ \bibnamefont {Schunck}}, \ and\ \bibinfo {author}
  {\bibfnamefont {W.}~\bibnamefont {Ketterle}},\ }\href {\doibase
  10.1103/PhysRevLett.89.203201} {\bibfield  {journal} {\bibinfo  {journal}
  {Phys. Rev. Lett.}\ }\textbf {\bibinfo {volume} {89}},\ \bibinfo {pages}
  {203201} (\bibinfo {year} {2002})}\BibitemShut {NoStop}%
\bibitem [{\citenamefont {Bourdel}\ \emph {et~al.}(2003)\citenamefont
  {Bourdel}, \citenamefont {Cubizolles}, \citenamefont {Khaykovich},
  \citenamefont {Magalh\~aes}, \citenamefont {Kokkelmans}, \citenamefont
  {Shlyapnikov},\ and\ \citenamefont {Salomon}}]{Bourdel2003mot}%
  \BibitemOpen
  \bibfield  {author} {\bibinfo {author} {\bibfnamefont {T.}~\bibnamefont
  {Bourdel}}, \bibinfo {author} {\bibfnamefont {J.}~\bibnamefont {Cubizolles}},
  \bibinfo {author} {\bibfnamefont {L.}~\bibnamefont {Khaykovich}}, \bibinfo
  {author} {\bibfnamefont {K.~M.~F.}\ \bibnamefont {Magalh\~aes}}, \bibinfo
  {author} {\bibfnamefont {S.~J. J. M.~F.}\ \bibnamefont {Kokkelmans}},
  \bibinfo {author} {\bibfnamefont {G.~V.}\ \bibnamefont {Shlyapnikov}}, \ and\
  \bibinfo {author} {\bibfnamefont {C.}~\bibnamefont {Salomon}},\ }\href
  {\doibase 10.1103/PhysRevLett.91.020402} {\bibfield  {journal} {\bibinfo
  {journal} {Phys. Rev. Lett.}\ }\textbf {\bibinfo {volume} {91}},\ \bibinfo
  {pages} {020402} (\bibinfo {year} {2003})}\BibitemShut {NoStop}%
\bibitem [{\citenamefont {Regal}\ \emph {et~al.}(2004)\citenamefont {Regal},
  \citenamefont {Greiner},\ and\ \citenamefont {Jin}}]{Regal2004lom}%
  \BibitemOpen
  \bibfield  {author} {\bibinfo {author} {\bibfnamefont {C.~A.}\ \bibnamefont
  {Regal}}, \bibinfo {author} {\bibfnamefont {M.}~\bibnamefont {Greiner}}, \
  and\ \bibinfo {author} {\bibfnamefont {D.~S.}\ \bibnamefont {Jin}},\ }\href
  {\doibase 10.1103/PhysRevLett.92.083201} {\bibfield  {journal} {\bibinfo
  {journal} {Phys. Rev. Lett.}\ }\textbf {\bibinfo {volume} {92}},\ \bibinfo
  {pages} {083201} (\bibinfo {year} {2004})}\BibitemShut {NoStop}%
\bibitem [{\citenamefont {Jochim}(2004)}]{Jochim2004thesis}%
  \BibitemOpen
  \bibfield  {author} {\bibinfo {author} {\bibfnamefont {S.}~\bibnamefont
  {Jochim}},\ }\href
  {http://www.ultracold.at/theses/thesis_selim_jochim/thesis_selim_jochim.pdf}
  {\bibfield  {journal} {\bibinfo  {journal} {{PhD Thesis,
  Leopold-Franzens-Universit{\"at}, Innsbruck}}\ } (\bibinfo {year}
  {2004})}\BibitemShut {NoStop}%
\bibitem [{\citenamefont {Schwenk}\ and\ \citenamefont
  {Pethick}(2005)}]{Schwenk2005rfg}%
  \BibitemOpen
  \bibfield  {author} {\bibinfo {author} {\bibfnamefont {A.}~\bibnamefont
  {Schwenk}}\ and\ \bibinfo {author} {\bibfnamefont {C.~J.}\ \bibnamefont
  {Pethick}},\ }\href {\doibase 10.1103/PhysRevLett.95.160401} {\bibfield
  {journal} {\bibinfo  {journal} {Phys. Rev. Lett.}\ }\textbf {\bibinfo
  {volume} {95}},\ \bibinfo {pages} {160401} (\bibinfo {year}
  {2005})}\BibitemShut {NoStop}%
\bibitem [{\citenamefont {Li}\ \emph {et~al.}(2018)\citenamefont {Li},
  \citenamefont {Tiesinga},\ and\ \citenamefont {Kotochigova}}]{Ming2018oqm}%
  \BibitemOpen
  \bibfield  {author} {\bibinfo {author} {\bibfnamefont {M.}~\bibnamefont
  {Li}}, \bibinfo {author} {\bibfnamefont {E.}~\bibnamefont {Tiesinga}}, \ and\
  \bibinfo {author} {\bibfnamefont {S.}~\bibnamefont {Kotochigova}},\ }\href
  {\doibase 10.1103/PhysRevA.97.053627} {\bibfield  {journal} {\bibinfo
  {journal} {Phys. Rev. A}\ }\textbf {\bibinfo {volume} {97}},\ \bibinfo
  {pages} {053627} (\bibinfo {year} {2018})}\BibitemShut {NoStop}%
\bibitem [{\citenamefont {K\"ohl}\ \emph {et~al.}(2005)\citenamefont {K\"ohl},
  \citenamefont {Moritz}, \citenamefont {St\"oferle}, \citenamefont
  {G\"unter},\ and\ \citenamefont {Esslinger}}]{Koehl2005fai}%
  \BibitemOpen
  \bibfield  {author} {\bibinfo {author} {\bibfnamefont {M.}~\bibnamefont
  {K\"ohl}}, \bibinfo {author} {\bibfnamefont {H.}~\bibnamefont {Moritz}},
  \bibinfo {author} {\bibfnamefont {T.}~\bibnamefont {St\"oferle}}, \bibinfo
  {author} {\bibfnamefont {K.}~\bibnamefont {G\"unter}}, \ and\ \bibinfo
  {author} {\bibfnamefont {T.}~\bibnamefont {Esslinger}},\ }\href {\doibase
  10.1103/PhysRevLett.94.080403} {\bibfield  {journal} {\bibinfo  {journal}
  {Phys. Rev. Lett.}\ }\textbf {\bibinfo {volume} {94}},\ \bibinfo {pages}
  {080403} (\bibinfo {year} {2005})}\BibitemShut {NoStop}%
\bibitem [{\citenamefont {Conway}\ and\ \citenamefont
  {Wybourne}(1963)}]{Conwaylle1963}%
  \BibitemOpen
  \bibfield  {author} {\bibinfo {author} {\bibfnamefont {J.~G.}\ \bibnamefont
  {Conway}}\ and\ \bibinfo {author} {\bibfnamefont {B.~G.}\ \bibnamefont
  {Wybourne}},\ }\href {\doibase 10.1103/PhysRev.130.2325} {\bibfield
  {journal} {\bibinfo  {journal} {Phys. Rev.}\ }\textbf {\bibinfo {volume}
  {130}},\ \bibinfo {pages} {2325} (\bibinfo {year} {1963})}\BibitemShut
  {NoStop}%
\bibitem [{\citenamefont {Smith}\ and\ \citenamefont
  {Unsworth}(1965)}]{Smith1965ths}%
  \BibitemOpen
  \bibfield  {author} {\bibinfo {author} {\bibfnamefont {K.~F.}\ \bibnamefont
  {Smith}}\ and\ \bibinfo {author} {\bibfnamefont {P.~J.}\ \bibnamefont
  {Unsworth}},\ }\href {http://stacks.iop.org/0370-1328/86/i=6/a=311}
  {\bibfield  {journal} {\bibinfo  {journal} {Proceedings of the Physical
  Society}\ }\textbf {\bibinfo {volume} {86}},\ \bibinfo {pages} {1249}
  (\bibinfo {year} {1965})}\BibitemShut {NoStop}%
\bibitem [{\citenamefont {Yan}\ \emph {et~al.}(2013)\citenamefont {Yan},
  \citenamefont {Moses}, \citenamefont {Gadway}, \citenamefont {Covey},
  \citenamefont {Hazzard}, \citenamefont {Rey}, \citenamefont {Jin},\ and\
  \citenamefont {Ye}}]{Yan2013ood}%
  \BibitemOpen
  \bibfield  {author} {\bibinfo {author} {\bibfnamefont {B.}~\bibnamefont
  {Yan}}, \bibinfo {author} {\bibfnamefont {S.~A.}\ \bibnamefont {Moses}},
  \bibinfo {author} {\bibfnamefont {B.}~\bibnamefont {Gadway}}, \bibinfo
  {author} {\bibfnamefont {J.~P.}\ \bibnamefont {Covey}}, \bibinfo {author}
  {\bibfnamefont {K.~R.~A.}\ \bibnamefont {Hazzard}}, \bibinfo {author}
  {\bibfnamefont {A.~M.}\ \bibnamefont {Rey}}, \bibinfo {author} {\bibfnamefont
  {D.~S.}\ \bibnamefont {Jin}}, \ and\ \bibinfo {author} {\bibfnamefont
  {J.}~\bibnamefont {Ye}},\ }\href {\doibase 10.1038/nature12483} {\bibfield
  {journal} {\bibinfo  {journal} {Nature}\ }\textbf {\bibinfo {volume} {501}},\
  \bibinfo {pages} {521} (\bibinfo {year} {2013})}\BibitemShut {NoStop}%
\bibitem [{\citenamefont {Baier~et al.}(2018)}]{Baier2017prep}%
  \BibitemOpen
  \bibfield  {author} {\bibinfo {author} {\bibfnamefont {S.}~\bibnamefont
  {Baier~et al.}},\ }\href@noop {} {\bibfield  {journal} {\bibinfo  {journal}
  {in preparation}\ } (\bibinfo {year} {2018})}\BibitemShut {NoStop}%
\bibitem [{\citenamefont {Becher}\ \emph {et~al.}(2018)\citenamefont {Becher},
  \citenamefont {Baier}, \citenamefont {Aikawa}, \citenamefont {Lepers},
  \citenamefont {Wyart}, \citenamefont {Dulieu},\ and\ \citenamefont
  {Ferlaino}}]{Becher2018apo}%
  \BibitemOpen
  \bibfield  {author} {\bibinfo {author} {\bibfnamefont {J.~H.}\ \bibnamefont
  {Becher}}, \bibinfo {author} {\bibfnamefont {S.}~\bibnamefont {Baier}},
  \bibinfo {author} {\bibfnamefont {K.}~\bibnamefont {Aikawa}}, \bibinfo
  {author} {\bibfnamefont {M.}~\bibnamefont {Lepers}}, \bibinfo {author}
  {\bibfnamefont {J.-F.}\ \bibnamefont {Wyart}}, \bibinfo {author}
  {\bibfnamefont {O.}~\bibnamefont {Dulieu}}, \ and\ \bibinfo {author}
  {\bibfnamefont {F.}~\bibnamefont {Ferlaino}},\ }\href {\doibase
  10.1103/PhysRevA.97.012509} {\bibfield  {journal} {\bibinfo  {journal} {Phys.
  Rev. A}\ }\textbf {\bibinfo {volume} {97}},\ \bibinfo {pages} {012509}
  (\bibinfo {year} {2018})}\BibitemShut {NoStop}%
\bibitem [{\citenamefont {J\"ordens}\ \emph {et~al.}(2008)\citenamefont
  {J\"ordens}, \citenamefont {Strohmaier}, \citenamefont {G\"unter},
  \citenamefont {Moritz},\ and\ \citenamefont {Essslinger}}]{Jordens2008ami}%
  \BibitemOpen
  \bibfield  {author} {\bibinfo {author} {\bibfnamefont {R.}~\bibnamefont
  {J\"ordens}}, \bibinfo {author} {\bibfnamefont {N.}~\bibnamefont
  {Strohmaier}}, \bibinfo {author} {\bibfnamefont {K.}~\bibnamefont
  {G\"unter}}, \bibinfo {author} {\bibfnamefont {H.}~\bibnamefont {Moritz}}, \
  and\ \bibinfo {author} {\bibfnamefont {T.}~\bibnamefont {Essslinger}},\
  }\href {\doibase 10.1038/nature07244} {\bibfield  {journal} {\bibinfo
  {journal} {Nature (London)}\ }\textbf {\bibinfo {volume} {455}},\ \bibinfo
  {pages} {204} (\bibinfo {year} {2008})}\BibitemShut {NoStop}%
\end{thebibliography}
\end{document}